\documentclass[twocolumn,english,aps,prl,showpacs,superscriptaddress,groupedaddress]{revtex4-1}
\usepackage[T1]{fontenc}
\usepackage[latin9]{inputenc}
\usepackage{color}
\usepackage{babel}
\usepackage{amsmath}
\usepackage{amssymb}
\usepackage{graphicx}
\usepackage[unicode=true,
 bookmarks=false,
 breaklinks=false,pdfborder={0 0 1},backref=section,colorlinks=false]
 {hyperref}
\usepackage{breakurl}

\makeatletter
\usepackage{babel}

\makeatother

\begin{document}

\title{Classification and Geometry of General Perceptual Manifolds}


\author{SueYeon Chung$^{1,2}$, Daniel D. Lee$^{2,3}$, and Haim Sompolinsky$^{2,4,5}$}

\affiliation{$^{1}$Program in Applied Physics, School of Engineering and Applied
Sciences, Harvard University, Cambridge, MA 02138, USA~~~~~~~\\
 $^{2}$Center for Brain Science, Harvard University, Cambridge, MA
02138, USA~~~~~~~\\
 $^{3}$School of Engineering and Applied Science, University of Pennysylvania,
Philadelphia, PA 19104, USA~~~~~~~\\
 $^{4}$Racah Institute of Physics, Hebrew University, Jerusalem 91904,
Israel~~~~~~~\\
 $^{5}$Edmond and Lily Safra Center for Brain Sciences, Hebrew University,
Jerusalem 91904, Israel}
\begin{abstract}
Perceptual manifolds arise when a neural population responds to an
ensemble of sensory signals associated with different physical features
(e.g., orientation, pose, scale, location, and intensity) of the same
perceptual object. Object recognition and discrimination require classifying
the manifolds in a manner that is insensitive to variability within
a manifold. How neuronal systems give rise to invariant object classification
and recognition is a fundamental problem in brain theory as well as
in machine learning. Here we study the ability of a readout network
to classify objects from their perceptual manifold representations.
We develop a statistical mechanical theory for the linear classification
of manifolds with arbitrary geometry revealing a remarkable relation
to the mathematics of conic decomposition. We show how special \emph{anchor
points} on the manifolds can be used to define novel geometrical measures
of \emph{radius} and \emph{dimension} which can explain the classification
capacity for manifolds of various geometries. The general theory is
demonstrated on a number of representative manifolds, including $\ell_{2}$
ellipsoids prototypical of strictly convex manifolds, $\ell_{1}$
balls representing polytopes with finite samples, and ring manifolds
exhibiting non-convex continuous structures that arise from modulating
a continuous degree of freedom. The effects of label sparsity on the
classification capacity of general manifolds are elucidated, displaying
a universal scaling relation between label sparsity and the manifold
radius. Theoretical predictions are corroborated by numerical simulations
using recently developed algorithms to compute maximum margin solutions
for manifold dichotomies. Our theory and its extensions provide a
powerful and rich framework for applying statistical mechanics of
linear classification to data arising from perceptual neuronal responses
as well as to artificial deep networks trained for object recognition
tasks. 
\end{abstract}
\maketitle

\section{Introduction\label{sec:Introduction}}

One fundamental cognitive task performed by animals and humans is
the invariant perception of objects, requiring the nervous system
to discriminate between different objects despite substantial variability
in each objects physical features. For example, in vision, the mammalian
brain is able to recognize objects despite variations in their orientation,
position, pose, lighting and background. Such impressive robustness
to physical changes is not limited to vision; other examples include
speech processing which requires the detection of phonemes despite
variability in the acoustic signals associated with individual phonemes;
and the discrimination of odors in the presence of variability in
odor concentrations. Sensory systems are organized as hierarchies
consisting of multiple layers transforming sensory signals into a
sequence of distinct neural representations. Studies of high level
sensory systems, e.g., the inferotemporal cortex (IT) in vision \citep{dicarlo2007untangling},
auditory cortex in audition \citep{bizley2013and}, and piriform cortex
in olfaction \citep{bolding2017complementary}, reveal that even the
late sensory stages exhibit significant sensitivity of neuronal responses
to physical variables. This suggests that sensory hierarchies generate
representations of objects that although not entirely invariant to
changes in physical features, are still readily decoded in an invariant
manner by a downstream system. This hypothesis is formalized by the
notion of the untangling of perceptual manifolds \citep{seung2000manifold,dicarlo2012does,poole2016exponential}.
This viewpoint underlies a number of studies of object recognition
in deep neural networks for artificial intelligence \citep{ranzato2007unsupervised,bengio2009learning,goodfellow2009measuring,cadieu2014deep}.

\begin{figure}[h]
\begin{centering}
\includegraphics[width=1\columnwidth]{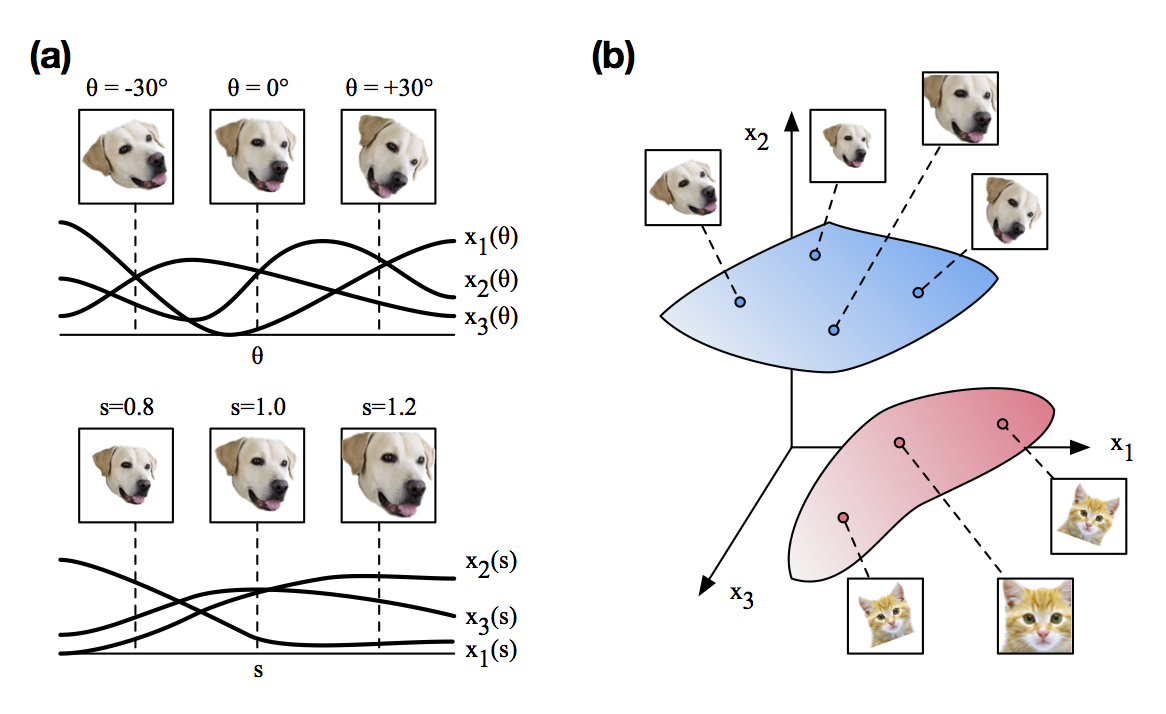} 
\par\end{centering}
\caption{Perceptual manifolds in neural state space.\textbf{ }(a) Firing rates
of neurons responding to images of a dog shown at various orientations
$\theta$ and scales $s$. The response to a particular orientation
and scale can be characterized by an $N$-dimensional population response.
(b) The population responses to the images of the dog form a continuous
manifold representing the complete set of invariances in the $\mathbb{R}^{N}$
neural activity space. Other object images, such as those corresponding
to a cat in various poses, are represented by other manifolds in this
vector space. \label{fig:Manifold_Illustration}}
\end{figure}

To conceptualize perceptual manifolds, consider a set of $N$ neurons
responding to a specific sensory signal associated with an object
as shown in Fig. \ref{fig:Manifold_Illustration}. The neural population
response to that stimulus is a vector in $\mathbb{R}^{N}$. Changes
in the physical parameters of the input stimulus that do not change
the object identity modulate the neural state vector. The set of all
state vectors corresponding to responses to all possible stimuli associated
with the same object can be viewed as a manifold in the neural state
space. In this geometrical perspective, object recognition is equivalent
to the task of discriminating manifolds of different objects from
each other. Presumably, as signals propagate from one processing stage
to the next in the sensory hierarchy, the geometry of the manifolds
is reformatted so that they become ``untangled,'' namely they are
more easily separated by a biologically plausible decoder \citep{dicarlo2007untangling}.
In this paper, we model the decoder as a simple single layer network
(the perceptron) and ask how the geometrical properties of the perceptual
manifolds influence their ability to be separated by a linear classifier.

Linear separability has previously been studied in the context of
the classification of points by a perceptron, using combinatorics
\citep{cover1965geometrical} and statistical mechanics \citep{gardner1988space,gardner87}.
Gardner's statistical mechanics theory is extremely important as it
provides accurate estimates of the perceptron capacity beyond function
counting by incorporating robustness measures. The robustness of a
linear classifier is quantified by the margin, which measures the
distance between the separating hyperplane and the closest point.
Maximizing the margin of a classifier is a critical objective in machine
learning, providing Support Vector Machines (SVM) with their good
generalization performance guarantees \citep{vapnik1998statistical}.

The above theories focus on separating a finite set of points with
no underlying geometrical structure and are not applicable to the
problem of manifold classification which deals with separating infinite
number of points geometrically organized as manifolds. This paper
addresses the important question of how to quantify the capacity of
the perceptron for dichotomies of input patterns described by manifolds.
In an earlier paper, we have presented the analysis for classification
of manifolds of extremely simple geometry, namely balls \citep{chung2016linear}.
However, the previous results have limited applicability as the neural
manifolds arising from realistic physical variations of objects can
exhibit much more complicated geometries. Can statistical mechanics
deal with the classification of manifolds with complex geometry, and
what specific geometric properties determine the separability of manifolds?

In this paper, we develop a theory of the linear separability of general,
finite dimensional manifolds. The summary of our key results is as
follows: 
\begin{itemize}
\item We begin by introducing a mathematical model of general manifolds
for binary classification (Sec. \ref{secManifold-Model}). This formalism
allows us to generate generic bounds on the manifold separability
capacity from the limits of small manifold sizes (classification of
isolated points) as that of large sizes (classification of entire
affine subspaces). These bounds highlight the fact that for large
ambient dimension $N$, the maximal number $P$ of separable finite-dimensional
manifolds is proportional to $N$, even though each consists of infinite
number of points, setting the stage for a statistical mechanical evaluation
of the maximal $\alpha=\frac{P}{N}$ . 
\item Using replica theory, we derive mean field equations for the capacity
of linear separation of finite dimensional manifolds (Sec. \ref{sec:Statistical-Mechanics})
and for the statistical properties of the optimal separating weight
vector. The mean-field solution is given in the form of self consistent
Karush-Kuhn-Tucker (KKT) conditions involving the \emph{manifold anchor
point}. The anchor point is a representative support vector for the
manifold. The position of the anchor point on a manifold changes as
the orientations of the other manifolds are varied, and the ensuing
statistics of the distribution of\emph{ anchor points} play a key
role in our theory. The optimal separating plane intersects a fraction
of the manifolds (the supporting manifolds). Our theory categorizes
the dimension of the span of the intersecting sets (e.g., points,
edges, faces, or full manifolds) in relation to the position of the
anchor points in the manifolds' convex hulls. 
\item The mean field theory motivates a new definition of manifold geometry,
which is based on the measure induced by the statistics of the anchor
points. In particular, we define the manifold anchor radius and dimension,
$R_{\text{M}}$ and $D_{\text{M}}$, respectively. These quantities
are relevant since the capacity of general manifolds can be well approximated
by the capacity of $\ell_{2}$ balls with radii $R_{\text{M}}$ and
dimensions $D_{\text{M}}$ . Interestingly, we show that in the limit
of small manifolds, the anchor point statistics are dominated by points
on the boundary of the manifolds which have minimal overlap with Gaussian
random vectors. The resultant Gaussian radius, $R_{\text{g}}$ and
dimension, $D_{\text{g}}$, are related to the well-known \emph{Gaussian
mean-width} of convex bodies (Sec. \ref{sec:Manifold-Geometry}).
Beyond understanding fundamental limits for classification capacity,
these geometric measures offer new quantitative tools for assessing
how perceptual manifolds are reformatted in brain and artificial systems. 
\item We apply the general theory to three examples, representing distinct
prototypical manifold classes. One class consists of manifolds with
strictly smooth convex hulls, which do not contain facets and are
exemplified by $\ell_{2}$ ellipsoids. Another class is that of convex
polytopes, which arise when the manifolds consists of a finite number
of data points, and are exemplified by $\ell_{1}$ ellipsoids. Finally,
\emph{ring} manifolds represent an intermediate class: smooth but
nonconvex manifolds. Ring manifolds are continuous nonlinear functions
of a single intrinsic variable, such as object orientation angle.
The differences between these manifold types show up most clearly
in the distinct patterns of the support dimensions. However, as we
show, they share common trends. As the size of the manifold increases,
the capacity and geometrical measures vary smoothly, exhibiting a
smooth cross-over from small radius and dimension with high capacity
to large radius and dimension with low capacity. This crossover occurs
as $R_{\text{g}}\propto\frac{1}{\sqrt{D_{\text{g}}}}$. Importantly,
for many realistic cases, when the size is smaller than the crossover
value, the manifold dimensionality is substantially smaller than that
computed from naive second order statistics, highlighting the saliency
and significance of our measures for the anchor geometry. 
\item Finally, we treat the important case of the classification of manifolds
with imbalanced (sparse) labels, which commonly arise in problems
of object recognition. It is well known that in highly sparse labels,
the classification capacity of random points increases dramatically
as $\left(f\left|\log f\right|\right)^{-1}$, where $f\ll1$ is the
fraction of the minority labels. Our analysis of sparsely labeled
manifolds highlights the interplay between manifold size and sparsity.
In particular, it shows that sparsity enhances the capacity only when
$fR_{\text{g}}^{2}\ll1,$ where $R_{\text{g}}$ is the (Gaussian)
manifold radius. Notably, for a large regime of parameters, sparsely
labeled manifolds can approximately be described by a universal capacity
function equivalent to sparsely labeled $\ell_{2}$ balls with radii
$R_{\text{g}}$ and dimensions $D_{\text{g}}$, as demonstrated by
our numerical evaluations (Sec. \ref{sec:Sparse-Labels}). Conversely,
when $fR_{\text{g}}^{2}\gg1$ , the capacity is low and close to $\frac{1}{D}$
where $D$ is the dimensionality of the manifold affine subspace,
even for extremely small $f$. 
\item Our theory provides for the first time, quantitative and qualitative
predictions for the perceptron classification of realistic data structures.
However, application to real data may require further extensions of
the theory and are discussed in Section \ref{sec:Summary}. Together,
the theory makes an important contribution to the development of statistical
mechanical theories of neural information processing in realistic
conditions. 
\end{itemize}

\section{Model of Manifolds\label{secManifold-Model}}

\begin{figure}
\begin{centering}
\includegraphics[width=1\columnwidth]{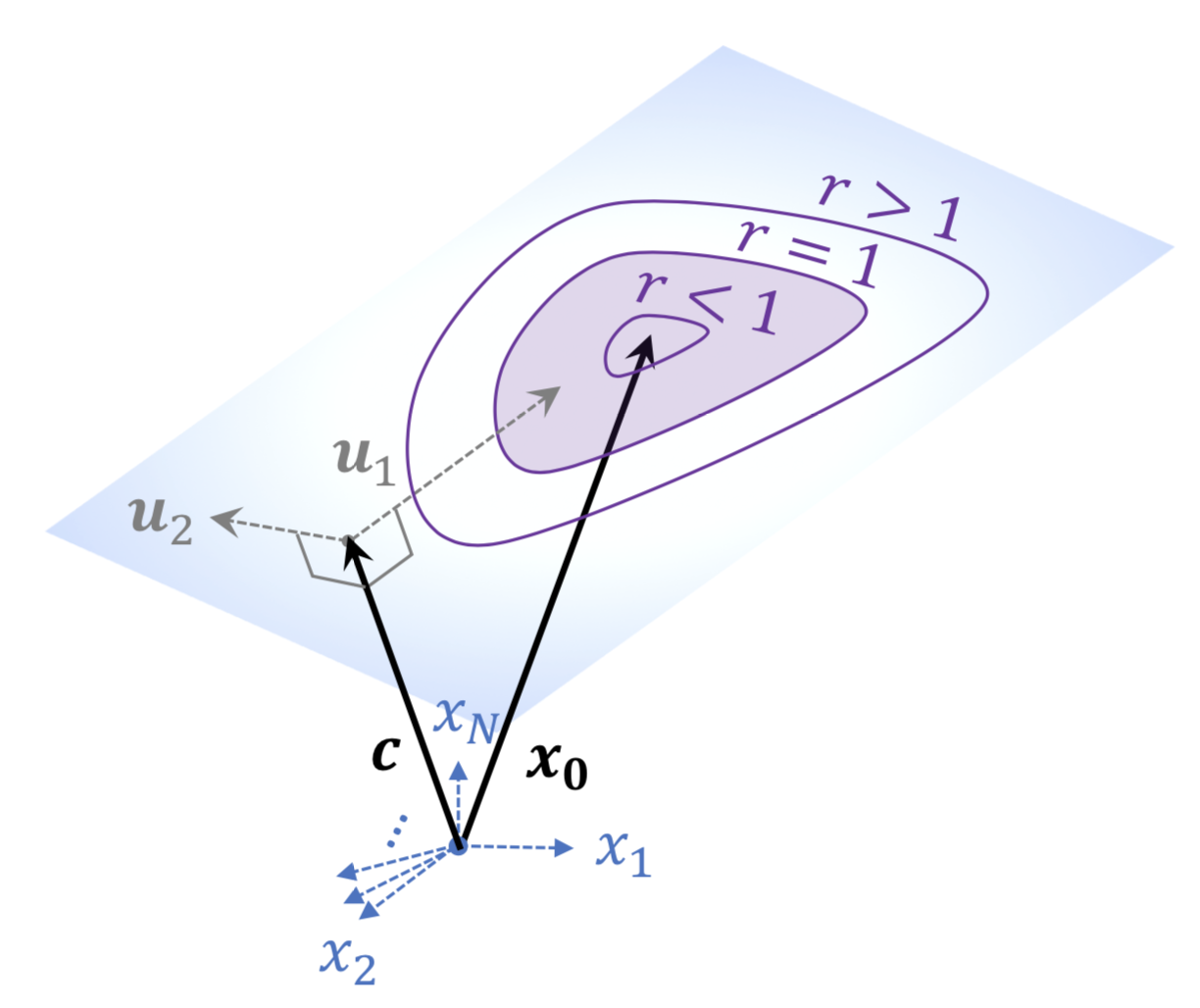} 
\par\end{centering}
\caption{Model of manifolds in affine subspaces. $D=2$ manifold embedded in
$\mathbb{R}^{N}$. $\mathbf{c}$ is the orthogonal translation vector
for the affine space and \textbf{$\mathbf{x_{0}}$} is the center
of the manifold. As the scale $r$ is varied, the manifold shrinks
to the point $\mathbf{x_{0}}$, or expands to fill the entire affine
space. \label{fig:Model-of-Manifolds}}
\end{figure}

\textbf{Manifolds in affine subspaces}: We model a set of $P$ perceptual
manifolds corresponding to $P$ perceptual object. Each manifold $M^{\mu}$
for $\mu=1,\ldots,P$ consists of a compact subset of an affine subspace
of $\mathbb{R}^{N}$ with \emph{affine dimension} $D$ with $D<N$.
A point on the manifold $\mathbf{x}^{\mu}\in M^{\mu}$ can be parameterized
as: 
\begin{equation}
\mathbf{x}^{\mu}(\vec{S})=\sum_{i=1}^{D+1}S_{i}\mathbf{u}_{i}^{\mu},\label{eq:d+1manifolds}
\end{equation}
where the $\mathbf{u}_{i}^{\mu}$ are a set of orthonormal bases of
the ($D+1$)-dimensional linear subspace containing $M^{\mu}$. The
$D+1$ components $S_{i}$ represents the coordinates of the manifold
point within this subspace and are constrained to be in the set $\vec{S}\in\mathcal{S}$.
The bold notation for $\mathbf{x}^{\mu}$ and $\mathbf{u}_{i}^{\mu}$
indicates they are vectors in $\mathbb{R}^{N}$ whereas the arrow
notation for $\vec{S}$ indicates it is a vector in $\mathbb{R}^{D+1}$.
The set $\mathcal{S}$ defines the shape of the manifolds and encapsulates
the affine constraint. For simplicity, we will first assume the manifolds
have the same geometry so that the coordinate set $\mathcal{S}$ is
the same for all the manifolds; extensions that consider heterogeneous
geometries are provided in Sec. \ref{subsec:Ensembles}.

We study the separability of $P$ manifolds into two classes, denoted
by binary labels $y^{\mu}=\pm1$, by a linear hyperplane passing through
the origin. A hyperplane is described by a weight vector $\mathbf{w}\in\mathbb{R}^{N}$,
normalized so $\left\Vert \mathbf{w}\right\Vert ^{2}=N$ and the hyperplane
correctly separates the manifolds with a margin $\kappa\ge0$ if it
satisfies, 
\begin{equation}
y^{\mu}\mathbf{w}\cdot\mathbf{x}^{\mu}\geq\kappa\label{eq:linearSep}
\end{equation}
for all $\mu$ and $\mathbf{x}^{\mu}\in M^{\mu}$. Since linear separability
is a convex problem, separating the manifolds is equivalent to separating
the convex hulls, $\text{conv}\left(M^{\mu}\right)=\left\{ \mathbf{x}^{\mu}(\vec{S})\mid\vec{S}\in\text{conv\ensuremath{\left(\mathcal{S}\right)}}\right\} $,
where 
\begin{equation}
\text{conv}\left(\mathcal{S}\right)=\left\{ \sum_{i=1}^{D+1}\alpha_{i}\vec{S}_{i}\mid\vec{S}_{i}\in{\cal S},\,\alpha_{i}\ge0,\,\sum_{i=1}^{D+1}\alpha_{i}=1\right\} .
\end{equation}

The position of an affine subspace relative to the origin can be defined
via the translation vector that is closest to the origin. This \emph{orthogonal
translation vector} $\mathbf{c}^{\mu}$ is perpendicular to all the
affine displacement vectors in $M^{\mu}$, so that all the points
in the affine subspace have equal projections on $\mathbf{c}^{\mu}$,
i.e., $\vec{\mathbf{x}}^{\mu}\cdot\mathbf{c}^{\mu}=\left\Vert \mathbf{c}^{\mu}\right\Vert ^{2}$
for all $\mathbf{x}^{\mu}\in M^{\mu}$ (Fig. \ref{fig:Model-of-Manifolds}).
We will further assume for simplicity that the $P$ norms$\left\Vert \mathbf{c}^{\mu}\right\Vert $
are all the same and normalized to $1$.

To investigate the separability properties of manifolds, it is helpful
to consider scaling a manifold $M^{\mu}$ by an overall scale factor
$r$ without changing its shape. We define the scaling relative to
a center $\vec{S}^{0}\in\mathcal{S}$ by a scalar $r>0$, by 
\begin{equation}
rM^{\mu}=\left\{ \sum_{i=1}^{D+1}\left[S_{i}^{0}+r\left(S_{i}-S_{i}^{0}\right)\right]\mathbf{u}_{i}^{\mu}\mid\vec{S}\in\mathcal{S}\right\} \label{eq:rMmu}
\end{equation}
When $r\rightarrow0$, the manifold $rM^{\mu}$ converges to a point:
$\mathbf{x}_{0}^{\mu}=\sum_{i}S_{i}^{0}\mathbf{u}_{i}^{\mu}$. On
the other hand, when $r\rightarrow\infty$, the manifold $rM^{\mu}$
spans the entire affine subspace. If the manifold is symmetric (such
as for an ellipsoid), there is a natural choice for a center. We will
later provide an appropriate definition for the center point for general,
asymmetric manifolds. In general, the translation vector $\vec{c}$
and center $\vec{S}^{0}$ need not coincide as shown in Fig. \ref{fig:Model-of-Manifolds}.
However, we will also discuss later the special case of \emph{centered}
manifolds in which the translation vector and center do coincide.

\textbf{Bounds on linear separability of manifolds:} For dichotomies
of $P$ input points in $\mathbb{R}^{N}$ at zero margin, $\kappa=0$,
the number of dichotomies that can be separated by a linear hyperplane
through the origin is given by \citep{cover1965geometrical}: 
\begin{equation}
C_{0}(P,N)=2\sum_{k=0}^{N-1}\text{C}_{k}^{P-1}\le2^{P}
\end{equation}
where $\text{C}_{k}^{n}=\frac{n!}{k!(n-k)!}$ is the binomial coefficient
for $n\ge k$, and zero otherwise. This result holds for $P$ input
vectors that obey the mild condition that the vectors are in \emph{general
position}, namely that all subsets of input vectors of size $p\le N$
are linearly independent.

For large $P$ and $N$, the probability $\frac{1}{2^{P}}C_{0}(P,N)$
of a dichotomy being linearly separable depends only upon the ratio
$\frac{P}{N}$ and exhibits a sharp transition at the critical value
of $\alpha_{0}=2$. We are not aware of a comprehensive extension
of Cover's counting theorem for general manifolds; nevertheless, we
can provide lower and upper bounds on the number of linearly realizable
dichotomies by considering the limit of $r\rightarrow0$ and $r\rightarrow\infty$
under the following general conditions.

First, in the limit of $r\rightarrow0$, the linear separability of
$P$ manifolds becomes equivalent to the separability of the $P$
centers. This leads to the requirement that the centers of the manifolds,
$\mathbf{x}_{0}^{\mu}$, are in general position in $\mathbb{R}^{N}$.
Second, we consider the conditions under which the manifolds are linearly
separable when $r\rightarrow\infty$ so that the manifolds span complete
affine subspaces. For a weight vector $\mathbf{w}$ to consistently
assign the same label to all points on an affine subspace, it must
be orthogonal to all the displacement vectors in the affine subspace.
Hence, to realize a dichotomy of $P$ manifolds when $r\rightarrow\infty$,
the weight vector $\mathbf{w}$ must lie in a null space of dimension
$N-D_{tot}$ where $D_{tot}$ is the rank of the union of affine displacement
vectors. When the basis vectors $\mathbf{u}_{i}^{\mu}$ are in general
position, then $D_{tot}=\min\left(DP,N\right)$. Then for the affine
subspaces to be separable, $PD<N$ is required and the projections
of the $P$ orthogonal translation vectors need also be separable
in the $N-D_{tot}$ dimensional null space. Under these general conditions,
the number of dichotomies for $D$-dimensional affine subspaces that
can be linearly separated, $C_{D}(P,N)$, can be related to the number
of dichotomies for a finite set of points via: 
\begin{equation}
C_{D}(P,N)=C_{0}(P,N-PD).
\end{equation}

From this relationship, we conclude that the ability to linearly separate
$D$-dimensional affine subspaces exhibits a transition from always
being separable to never being separable at the critical ratio $\frac{P}{N}=\frac{2}{1+2D}$
for large $P$ and $N$ (see Supplementary Materials, SM, Sec. S1).

For general $D$-dimensional manifolds with finite size, the number
of dichotomies that are linearly separable will be lower bounded by
$C_{D}(P,N)$ and upper bounded by $C_{0}(P,N)$. We introduce the
notation, $\alpha_{\text{M}}(\kappa)$, to denote the maximal load
$\frac{P}{N}$ such that randomly labeled manifolds are linearly separable
with a margin $\kappa$, with high probability. Therefore, from the
above considerations, it follows that the critical load at zero margin,
$\alpha_{\text{M}}(\kappa=0)$, is bounded by, 
\begin{equation}
\frac{1}{2}\leq\alpha_{\text{M}}^{-1}(\kappa=0)\leq\frac{1}{2}+D.\label{eq:PN bound}
\end{equation}

These bounds highlight the fact that in the large $N$ limit, the
maximal number of separable finite-dimensional manifolds is proportional
to $N$, even though each consists of an infinite number of points.
This sets the stage of a statistical mechanical evaluation of the
maximal $\alpha=\frac{P}{N}$ , where $P$ is the number of manifolds,
and is described in the following Section.

\section{Statistical Mechanical Theory\label{sec:Statistical-Mechanics}}

In order to make theoretical progress beyond the bounds above, we
need to make additional statistical assumptions about the manifold
spaces and labels. Specifically, we will assume that the individual
components of $\mathbf{u}_{i}^{\mu}$ are drawn independently and
from identical Gaussian distributions with zero mean and variance
$\frac{1}{N}$, and that the binary labels $y^{\mu}=\pm1$ are randomly
assigned to each manifold with equal probabilities. We will study
the thermodynamic limit where $N,P\rightarrow\infty$, but with a
finite load $\alpha=\frac{P}{N}$. In addition, the manifold geometries
as specified by the set $\mathcal{S}$ in $\mathbb{R}^{D+1}$, and
in particular their affine dimension, $D$, is held fixed in the thermodynamic
limit. Under these assumptions, the bounds in Eq. \eqref{eq:PN bound}
can be extended to the linear separability of general manifolds with
finite margin $\kappa$, and characterized by the reciprocal of the
critical load ratio $\alpha_{\text{M}}^{-1}(\kappa)$, 
\begin{equation}
\alpha_{0}^{-1}(\kappa)\le\alpha_{\text{M}}^{-1}(\kappa)\le\alpha_{0}^{-1}(\kappa)+D\label{eq:PNboundGardner}
\end{equation}
where $\alpha_{0}(\kappa)$ is the maximum load for separation of
random i.i.d. points with a margin $\kappa$ given by the Gardner
theory \citep{gardner1988space}, 
\begin{equation}
\alpha_{0}^{-1}(\kappa)=\int_{-\infty}^{\kappa}Dt\,(t-\kappa)^{2}\label{eq:alpha0}
\end{equation}
with Gaussian measure $Dt=\frac{1}{\sqrt{2\pi}}e^{-\frac{t^{2}}{2}}$.
For many interesting cases, the affine dimension $D$ is large and
the gap in Eq. \eqref{eq:PNboundGardner} is overly loose. Hence,
it is important to derive an estimate of the capacity for manifolds
with finite sizes and evaluate the dependence of the capacity and
the nature of the solution on the geometrical properties of the manifolds
as shown below.

\subsection{Mean field theory of manifold separation capacity}

Following Gardner's framework \citep{gardner1988space,gardner87},
we compute the statistical average of $\text{\ensuremath{\log}}\,Z$,
where $Z$ is the volume of the space of the solutions, which in our
case can be written as: 
\begin{equation}
Z=\int d^{N}\mathbf{w}\delta\left(\left\Vert \mathbf{w}\right\Vert ^{2}-N\right)\,\Pi_{\mu,\mathbf{x}^{\mu}\in M^{\mu}}\Theta\left(y^{\mu}\mathbf{w}\cdot\mathbf{x}^{\mu}-\kappa\right),\label{eq:V}
\end{equation}
$\Theta\left(\cdot\right)$ is the Heaviside function to enforce the
margin constraints in Eq. \eqref{eq:linearSep}, along with the delta
function to ensure $\left\Vert \mathbf{w}\right\Vert ^{2}=N$. In
the following, we focus on the properties of the maximum margin solution,
namely the solution for the largest load $\alpha_{\text{M}}$ for
a fixed margin $\kappa$, or equivalently, the solution when the margin
$\kappa$ is maximized for a given $\alpha_{\text{M}}$.

As shown in Appendix A, we prove that the general form of the inverse
capacity, exact in the thermodynamic limit, is: 
\begin{equation}
\alpha_{\text{M}}^{-1}(\kappa)=\langle F(\vec{T})\rangle_{\vec{T}}\label{eq:inv_capacity}
\end{equation}
where $F(\vec{T})=\min_{\vec{V}}\left\{ \left\Vert \vec{V}-\vec{T}\right\Vert ^{2}\mid\,\vec{V}\cdot\vec{S}-\kappa\geq0,\,\forall\vec{S}\in\mathcal{S}\right\} $
and $\left\langle \ldots\right\rangle _{\vec{T}}$ is an average over
random $D+1$ dimensional vectors $\vec{T}$ whose components are
i.i.d. normally distributed $T_{i}\sim\mathcal{N}(0,1)$. The components
of the vector $\vec{V}$ represent the signed fields induced by the
solution vector $\mathbf{w}$ on the $D+1$ basis vectors of the manifold.
The Gaussian vector $\vec{T}$ represents the part of the variability
in $\vec{V}$ due to \emph{quenched} variability in the manifolds
basis vectors and the labels, as will be explained in detail below.

The inequality constraints in $F$ can be written equivalently, as
a constraint on the point on the manifold with \emph{minimal} projection
on $\vec{V}$. We therefore consider the concave\emph{ support function}
of $\mathcal{S}$, $g_{\mathcal{S}}(\vec{V})=\min_{\vec{S}}\left\{ \vec{V}\cdot\vec{S}\mid\vec{S}\in{\cal \mathcal{S}}\right\} $,
which can be used to write the constraint for $F(\vec{T})$ as

\begin{equation}
F(\vec{T})=\min_{\vec{V}}\left\{ \left\Vert \vec{V}-\vec{T}\right\Vert ^{2}\mid g_{\mathcal{S}}(\vec{V})-\kappa\geq0\right\} \label{eq:kappaG}
\end{equation}
Note that this definition of $g_{\mathcal{S}}(\vec{V})$ is easily
mapped to the conventional convex support function defined via the
\emph{max} operation\citep{boyd2004convex}.

\textbf{Karush-Kuhn-Tucker (KKT) conditions:} To gain insight into
the nature of the maximum margin solution, it is useful to consider
the KKT conditions of the convex optimization in Eq. \ref{eq:kappaG}
\citep{boyd2004convex}. For each $\vec{T}$, the KKT conditions that
characterize the unique solution of $\vec{V}$ for $F(\vec{T}$) is
given by:

\begin{equation}
\vec{V}=\vec{T}+\lambda\tilde{S}(\vec{T})\label{eq:vlambdagrad}
\end{equation}
where 
\begin{align}
\lambda & \geq0\label{eq:KKT}\\
g_{\mathcal{S}}(\vec{V})-\kappa & \geq0\nonumber \\
\lambda\left[g_{\mathcal{S}}(\vec{V})-\kappa\right] & =0\nonumber 
\end{align}

The vector $\tilde{S}(\vec{T})$ is a \emph{subgradient} of the support
function at $\vec{V}$, $\tilde{S}(\vec{T})\in\partial g_{S}\left(\vec{V}\right)$
\citep{boyd2004convex}, i.e., it is a point on the convex hull of
$S$ with minimal overlap with $\vec{V}$. When the support function
is differentiable, the subgradient $\partial g_{S}\left(\vec{V}\right)$
is unique and is equivalent to the \emph{gradient} of the support
function, 
\begin{equation}
\tilde{S}(\vec{T})=\nabla g_{\mathcal{S}}(\vec{V})=\arg\min_{\vec{S}\in\mathcal{S}}\vec{V}\cdot\vec{S}\label{eq:support_gradient}
\end{equation}

Since the support function is positively homogeneous, $g_{\mathcal{S}}(\gamma\vec{V})=\gamma g_{\mathcal{S}}(\vec{V})$
for all $\gamma>0$; thus, $\nabla g_{\mathcal{S}}(\vec{V})$ depends
only on the unit vector $\hat{V}$. For values of $\vec{V}$ such
that $g_{\mathcal{S}}(\vec{V})$ is not differentiable, the subgradient
is not unique, but $\tilde{S}(\vec{T})$ is defined uniquely as the
particular subgradient that obeys the KKT conditions, Eq. \ref{eq:vlambdagrad}-\ref{eq:KKT}.
In the latter case, $\tilde{S}(\vec{T})\in\text{conv}\left(\mathcal{S}\right)$
may not reside in $S$ itself.

From Eq. \eqref{eq:vlambdagrad}, we see that the capacity can be
written in terms of $\tilde{S}(\vec{T})$ as, 
\begin{equation}
F(\vec{T})=\left\Vert \lambda\tilde{S}(\vec{T})\right\Vert ^{2}.\label{eq:flambdas}
\end{equation}

The scale factor $\lambda$ is either zero or positive, corresponding
to whether $g_{\mathcal{S}}(\vec{V})-\kappa$ is positive or zero.
If $g_{\mathcal{S}}(\vec{V})-\kappa$ is positive , then $\lambda=0$
meaning that $\vec{V}=\vec{T}$ and $\vec{T}\cdot\tilde{S}(\vec{T})-\kappa>0$.
If $\vec{T}\cdot\tilde{S}(\vec{T})-\kappa<0$ , then, $\lambda>0$
and $\vec{V}\neq\vec{T}$. In this case, multiplying \ref{eq:vlambdagrad}
with $\tilde{S}(\vec{T})$ yields $\lambda\left\Vert \tilde{S}(\vec{T})\right\Vert ^{2}=\vec{T}\cdot\tilde{S}(\vec{T})-\kappa$.
Thus, $\lambda$ obeys the self consistent equation,

\begin{equation}
\lambda=\frac{\left[-\vec{T}\cdot\tilde{S}(\vec{T})+\kappa\right]_{+}}{\left\Vert \tilde{S}(\vec{T})\right\Vert ^{2}}\label{eq:lambdaRectified}
\end{equation}
where the function $\left[x\right]_{+}=\max(x,0)$.

\subsection{Mean field interpretation of the KKT relations}

The KKT relations have a nice interpretation within the framework
of the mean field theory. The maximum margin solution vector can always
be written as a linear combination of a set of support vectors. Although
there are infinite numbers of input points in each manifold, the solution
vector can be decomposed into $P$ vectors, one per manifold,

\begin{equation}
\mathbf{w}=\sum_{\mu=1}^{P}\lambda_{\mu}y^{\mu}\mathbf{\tilde{x}}^{\mu},\,\lambda_{\mu}\geq0\label{eq:w expand}
\end{equation}

where $\mathbf{\tilde{x}}^{\mu}\in\text{conv}\left(M^{\mu}\right)$
is a vector in the convex hull of the $\mu$-th manifold. In the large
$N$ limit, these vectors are uncorrelated with each other. Hence,
squaring this equation, ignoring correlations between the different
contributions yields, $\left\Vert \mathbf{w}\right\Vert ^{2}=N=\sum_{\mu=1}^{P}\lambda_{\mu}^{2}\left\Vert \tilde{S}^{\mu}\right\Vert ^{2}$,
where $\tilde{S}^{\mu}$ are the coordinates of $\mathbf{\tilde{x}}^{\mu}$
in the $\mu$-th affine subspace of the $\mu$-th manifold, see Eq.
\eqref{eq:d+1manifolds} (Note: using the 'cavity' method, one can
show that the $\lambda$'s appearing in the mean field KKT equations
and the $\lambda$'s appearing in Eq. \ref{eq:w expand} differ by
an overall global scaling factor which accounts for the presence of
correlations between the individual terms in Eq. \ref{eq:w expand},
which we have neglected here for brevity; see \cite{gerl1994storage,kinzel1991dynamics}).
From this equation, it follows that $\alpha^{-1}=\frac{N}{P}=\left\langle \lambda^{2}\left\Vert \tilde{S}\right\Vert ^{2}\right\rangle $
which yields the KKT expression for the capacity, see Eqs. \ref{eq:inv_capacity}
and \ref{eq:flambdas}.

The KKT relations above, are self-consistent equations for the statistics
of $\lambda_{\mu}$and $\tilde{S}^{\mu}$. The mean field theory derives
the appropriate statistics from self-consistent equations of the fields
on a \emph{single} manifold. To see this, consider projecting the
solution vector $\mathbf{w}$ onto the affine subspace of one of the
manifolds, say $\mu=1$ . We define a $D+1$ dimensional vector, $\vec{V}^{1}$,
as $V_{i}^{1}=y^{1}\mathbf{w}\cdot\mathbf{u}_{i}^{1}$, $i=1,...,D+1$,
which are the signed fields of the solution on the affine basis vectors
of the $\mu=1$ manifold. Then, Eq. \eqref{eq:w expand} reduces to
$\vec{V}^{1}=\lambda_{1}$$\tilde{S}^{1}+\vec{T}$ where $\vec{T}$
represents the contribution from all the other manifolds and since
their subspaces are randomly oriented, this contribution is well described
as a random Gaussian vector. Finally, self consistency requires that
for fixed $\vec{T}$, $\tilde{S}^{1}$ is such that it has a minimal
overlap with $\vec{V}^{1}$ and represents a point residing on the
margin hyperplane, otherwise it will not contribute to the max margin
solution. Thus Eq. \ref{eq:vlambdagrad} is just the decomposition
of the field induced on a specific manifold into the contribution
induced by that specific manifold along with the contributions coming
from all the other manifolds. The self consistent equations \ref{eq:KKT}
as well as \ref{eq:lambdaRectified} relating $\lambda$ to the Gaussian
statistics of $\vec{T}$ then naturally follow from the requirement
that $\tilde{S}^{1}$ represents a support vector.

\subsection{Anchor points and manifold supports}

The vectors $\mathbf{\tilde{x}}^{\mu}$ contributing to the solution,
Eq. \eqref{eq:w expand}, play a key role in the theory. We will denote
them or equivalently their affine subspace components, $\tilde{S}^{\mu}$,\emph{
}as the\emph{ manifold anchor} \emph{points}. For a particular configuration
of manifolds, the manifolds could be replaced by an equivalent set
of $P$ anchor points to yield the same maximum margin solution. It
is important to stress, however, that an individual anchor point is
determined not only by the configuration of its associated manifold
but also by the random orientations of all the other manifolds. For
a fixed manifold, the location of its anchor point will vary with
the relative configurations of the other manifolds. This variation
is captured in mean field theory by the dependence of the anchor point
$\tilde{S}$ on the random Gaussian vector $\vec{T}$.

In particular, the position of the anchor point in the convex hull
of its manifold reflects the nature of the relation between that manifold
and the margin planes. In general, a fraction of the manifolds will
intersect with the margin hyperplanes, i.e., they have non-zero $\lambda$;
these manifolds are the support manifolds of the system. The nature
of this support varies and can be characterized by the dimension,
$k$, of the span of the intersecting set of $\text{conv}\left(\mathcal{S}\right)$
with the margin planes. Some support manifolds, which we call \emph{touching}
manifolds, intersect with the margin hyperplane only with their anchor
point. They have a support dimension $k=1$, and their anchor point
$\tilde{S}$ is on the boundary of $\mathcal{S}$. The other extreme
is \emph{fully} \emph{supporting} manifolds which completely reside
in the margin hyperplane. They are characterized by $k=D+1$. In this
case, $\vec{V}$ is parallel to the translation vector $\vec{c}$
of $S$. Hence, all the points in $S$ are support vectors, and all
have the same overlap, $\kappa$, with $\vec{V}$. The anchor point,
in this case, is the unique point in the interior of $\text{conv}(\mathcal{S})$
which obeys the self consistent equation, \ref{eq:vlambdagrad}, namely
that it balances the contribution from the other manifolds to zero
out the components of $\vec{V}$ that are orthogonal to $\vec{c}$.
In the case of smooth convex hulls (i.e., when $S$ is strongly convex),
no other manifold support configurations exist. For other types of
manifolds, there are also \emph{partially supporting} manifolds, whose
convex hull intersection with the margin hyperplanes consist of $k$
dimensional faces with $1<k<D+1$. The associated anchor points then
reside inside the intersecting face. For instance, $k=2$ implies
that $\tilde{S}$ lies on an edge whereas $k=3$ implies that $\tilde{S}$
lies on a planar 2-face of the convex hull. Determining the dimension
of the support structure that arises for various $\vec{T}$ is explained
below.

\subsection{Conic decomposition}

\begin{figure}
\begin{centering}
\includegraphics[width=0.9\columnwidth]{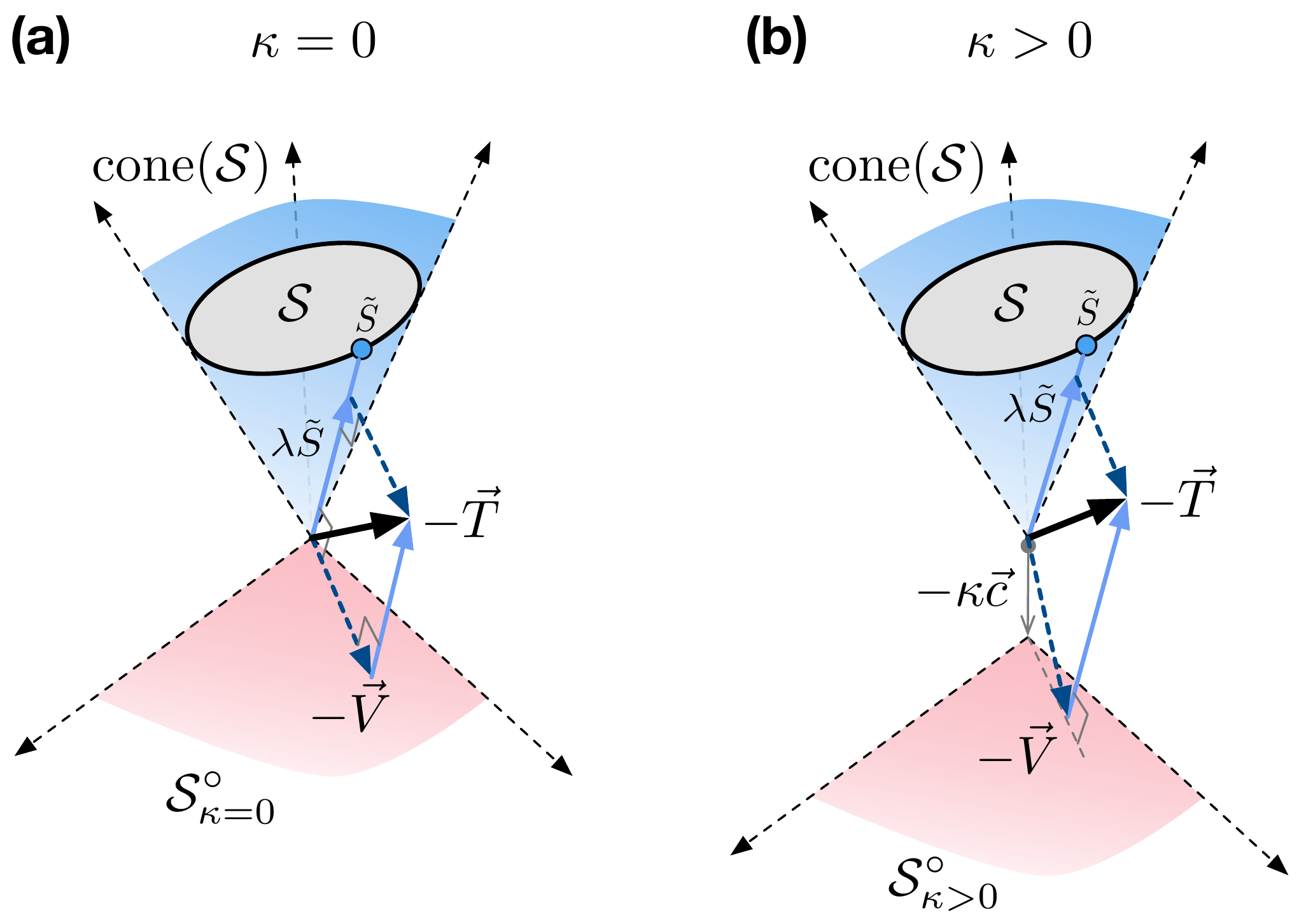} 
\par\end{centering}
\caption{Conic decomposition of $-\vec{T}=-\vec{V}+\lambda\tilde{S}$ at (a)
zero margin $\kappa=0$ and (b) non-zero margin $\kappa>0$. Given
a random Gaussian $\vec{T}$, $\vec{V}$ is found such that $\left\Vert \vec{V}-\vec{T}\right\Vert $
is minimized while $-\vec{V}$ being on the polar cone $\mathcal{S_{\kappa}^{\circ}}$.
$\lambda\tilde{S}$ is in $\text{cone}(\mathcal{S})$ and $\tilde{S}$
is the anchor point, a projection on the convex hull of $\mathcal{S}$.
\label{fig:ConicDecomposition}}
\end{figure}

The KKT conditions can also be interpreted in terms of the \emph{conic}
\emph{decomposition} of $\vec{T}$, which generalizes the notion of
the decomposition of vectors onto linear subspaces and their null
spaces via Euclidean projection. The \emph{convex cone} of the manifold
${\cal S}$ is defined as $\text{cone}\left(\mathcal{S}\right)=\left\{ \sum_{i=1}^{D+1}\alpha_{i}\vec{S}_{i}\mid\vec{S}_{i}\in{\cal S},\,\alpha_{i}\ge0\right\} $,
see Fig. \ref{fig:ConicDecomposition}. The \emph{shifted polar cone}
of ${\cal S}$, denoted $\mathcal{S}_{\kappa}^{\circ}$, is defined
as the convex set of points given by, 
\begin{equation}
{\cal S}_{\kappa}^{\circ}=\left\{ \vec{U}\in\mathbb{R}^{D+1}\mid\vec{U}\cdot\vec{S}+\kappa\le0,\:\forall\vec{S}\in{\cal S}\right\} \label{eq:shiftpolar}
\end{equation}
and is illustrated for $\kappa=0$ and $\kappa>0$ in Fig. \ref{fig:ConicDecomposition}.
For $\kappa=0$, Eq. \eqref{eq:shiftpolar} is simply the conventional
\emph{polar cone} of $\mathcal{S}$ \citep{rockafellar2015convex}.
Equation \eqref{eq:vlambdagrad} can then be interpreted as the decomposition
of $-\vec{T}$ into the sum of two component vectors: one component
is $-\vec{V}$, i.e., its Euclidean projection onto $\mathcal{S}_{\kappa}^{\circ}$;
the other component $\lambda\tilde{S}(\vec{T})$ is located in $\text{cone}\left(\mathcal{S}\right)$.
When $\kappa=0$, the Moreau decomposition theorem states that the
two components are perpendicular: $\vec{V}\cdot\left(\lambda\tilde{S}(\vec{T})\right)=0$
\citep{moreau1962decomposition}. For non-zero $\kappa$, the two
components need not be perpendicular but obey $\vec{V}\cdot\left(\lambda\tilde{S}(\vec{T})\right)=\kappa\lambda$.

The position of vector $\vec{T}$ in relation to the cones, $\text{cone}(\mathcal{S})$
and $\mathcal{S}_{\kappa}^{\circ}$, gives rise to qualitatively different
expressions for $F\left(\vec{T}\right)$ and contributions to the
solution weight vector and inverse capacity. These correspond to the
different support dimensions mentioned above. In particular, when
-$\vec{T}$ lies inside ${\cal S}_{\kappa}^{\circ}$ , $\vec{T}=\vec{V}$
and $\lambda=0$ so the support dimension, $k=0$. On the other hand,
when $-\vec{T}$ lies inside $\text{cone}\left(\mathcal{S}\right)$,
$\vec{V}=\kappa\vec{c}$ and the manifold is fully supporting, $k=D+1$.

\subsection{Numerical solution of the mean field equations \label{subsec:NumericalMFT}}

The solution of the mean field equations consists of two stages. First,
$\tilde{S}$ is computed for a particular $\vec{T}$ and then the
relevant contributions to the inverse capacity are averaged over the
Gaussian distribution of $\vec{T}$. For simple geometries such as
$\ell_{2}$ ellipsoids, the first step may be solved analytically.
However, for more complicated geometries, both steps need to be performed
numerically. The first step involves determining $\vec{V}$ and $\tilde{S}$
for a given $\vec{T}$ by solving the quadratic semi-infinite programming
problem (QSIP), Eq. \eqref{eq:kappaG}, over the manifold $\mathcal{S}$
which may contain infinitely many points. A novel ``cutting plane''
method has been developed to efficiently solve the QSIP problem, see
SM (Sec. S4). Expectations are computed by sampling Gaussian $\vec{T}$
in $D+1$ dimensions and taking the appropriate averages, similar
to procedures for other mean field methods. The relevant quantities
corresponding to the capacity are quite concentrated and converge
quickly with relatively few samples.

In following Sections, we will also show how the mean field theory
compares with computer simulations that numerically solve for the
maximum margin solution of realizations of $P$ manifolds in $\mathbb{R}^{N}$,
given by Eq. \eqref{eq:linearSep}, for a variety of manifold geometries.
Finding the maximum margin solution is challenging as standard methods
to solving SVM problems are limited to a finite number of input points.
We have recently developed an efficient algorithm for finding the
maximum margin solution in manifold classification and have used this
method in the present work (see \citep{chung2017learning} and SM
(Sec. S5)).

\section{Manifold Geometry\label{sec:Manifold-Geometry}}

\subsection{Longitudinal and intrinsic coordinates\label{subsec:Longitudianl-and-intrinsic}}

In this section, we address how the capacity to separate a set of
manifolds can be related to their geometry, in particular to their
shape within the $D$-dimensional affine subspace. Since the projections
of all points in a manifold onto the translation vector, $\mathbf{c}^{\mu}$,
are the same, $\vec{\mathbf{x}}^{\mu}\cdot\mathbf{c}^{\mu}=\left\Vert \mathbf{c}^{\mu}\right\Vert =1$,
it is convenient to parameterize the $D+1$ affine basis vectors such
that $\mathbf{u}_{D+1}^{\mu}=\mathbf{c}^{\mu}$. In these coordinates,
the $D+1$ dimensional vector representation of $\mathbf{c}^{\mu}$
is $\vec{C}=\left(0,0,\ldots,0,1\right)$. This parameterization is
convenient since it constrains the manifold variability to be in the
first $D$ components of $\vec{S}$ while the $D+1$ coordinate is
a longitudinal variable measuring the distance of the manifold affine
subspace from the origin. We can then write the $D+1$ dimensional
vectors $\vec{T}=(\vec{t},\,t_{0})$ where $t_{0}=\vec{T}\cdot\vec{C}$,
$\vec{V}=(\vec{v},\:v_{0}),$ $\vec{S}=$$(\vec{s},\,1)$, $\tilde{S}=(\tilde{s},\,1)$
, where lower case vectors denote vectors in $\mathbb{R}^{D}$. We
will also refer to the $D$-dimensional intrinsic vector, $\tilde{s}$,
as the anchor point. In this notation, the capacity, Eq. \eqref{eq:flambdas}-\eqref{eq:lambdaRectified}
can be written as, 
\begin{equation}
\alpha_{\text{M}}^{-1}(\kappa)=\int D\vec{t}\int Dt_{0}\frac{\left[-t_{0}-\vec{t}\cdot\tilde{s}(\vec{t},t_{0})+\kappa\right]_{+}^{2}}{1+\left\Vert \tilde{s}(\vec{t},t_{0})\right\Vert ^{2}}\label{eq:amMgen}
\end{equation}

It is clear from this form that when $D=0$, or when all the vectors
in the affine subspace obey $\tilde{s}(\vec{t},t_{0})=0$ , the capacity
reduces to the Gardner result, Eq. \eqref{eq:alpha0}. Since $\vec{V}\cdot\tilde{S}$=$\vec{v}\cdot\tilde{s}+v_{0},$
for all $\tilde{S}$ , $g(\vec{V})=g(\vec{v})+v_{0}$ , and the support
function can be expressed as, 
\begin{equation}
g(\vec{v})=\min_{\vec{s}}\left\{ \vec{v}\cdot\vec{s}\mid\vec{s}\in{\cal \mathcal{S}}\right\} .\label{eq:gsperp}
\end{equation}
The KKT relations can be written as $\vec{v}=\vec{t}+\lambda\vec{s}$
where $\lambda=v_{0}-t_{0}\geq0,$ $g(\vec{v})\geq\kappa-v_{0}$,
$\lambda\left[g(\vec{v})+v_{0}-\kappa\right]=0$, and $\tilde{s}$
minimizes the overlap with $\vec{v}$. The resultant equation for
$\lambda$ (or $v_{0})$ is $\lambda=$$\left[-t_{0}-\vec{t}\cdot\tilde{s}(\vec{t},t_{0})+\kappa\right]_{+}/\left(1+\left\Vert \tilde{s}(\vec{t},t_{0})\right\Vert ^{2}\right)$
which agrees with Eq. \ref{eq:lambdaRectified}.

\subsection{Types of supports \label{subsec:TypesSupport }}

Using the above coordinates, we can elucidate the conditions for the
different types of support manifolds defined in the previous section.
To do this, we fix the random vector $\vec{t}$ and consider the qualitative
change in the anchor point, $\tilde{s}(\vec{t},t_{0})$ as $t_{0}$
decreases from $+\infty$ to $-\infty$.

\textbf{Interior manifolds ($k=0$):} For sufficiently positive $t_{0}$,
the manifold is interior to the margin plane, i.e., $\lambda=0$ with
corresponding support dimension $k=0$. Although not contributing
to the inverse capacity and solution vector, it is useful to associate
anchor points to these manifolds defined as the closest point on the
manifold to the margin plane: $\tilde{s}(\vec{t})=\arg\min_{\vec{s}\in\mathcal{S}}\vec{t}\cdot\vec{s}=\nabla g(\vec{t})$
since $\vec{v}=\vec{t}$. This definition ensures continuity of the
anchor point for all $\lambda\ge0$.

This interior regime holds when $g(\vec{t})>\kappa-t_{0}$, or equivalently
for $t_{0}-\kappa>t_{\text{touch}}(\vec{t})$ where 
\begin{equation}
t_{\text{touch}}(\vec{t})=-g(\vec{t})\label{eq:tTouch}
\end{equation}

Non-zero contributions to the capacity only occur outside this interior
regime when $g(\vec{t})+t_{0}-\kappa<0$, in which case $\lambda>0$.
Thus, for all support dimensions $k>0$, the solution for $\vec{v}$
is \emph{active}, satisfying the equality condition, $g(\vec{v})+v_{0}-\kappa=0$,
so that from Eq. \eqref{eq:vlambdagrad}: 
\begin{equation}
g(\vec{v})+t_{0}-\kappa+\lambda=0,\label{eq:gKappaActive}
\end{equation}

outside the interior regime.

\begin{figure}
\begin{centering}
\includegraphics[width=0.8\columnwidth]{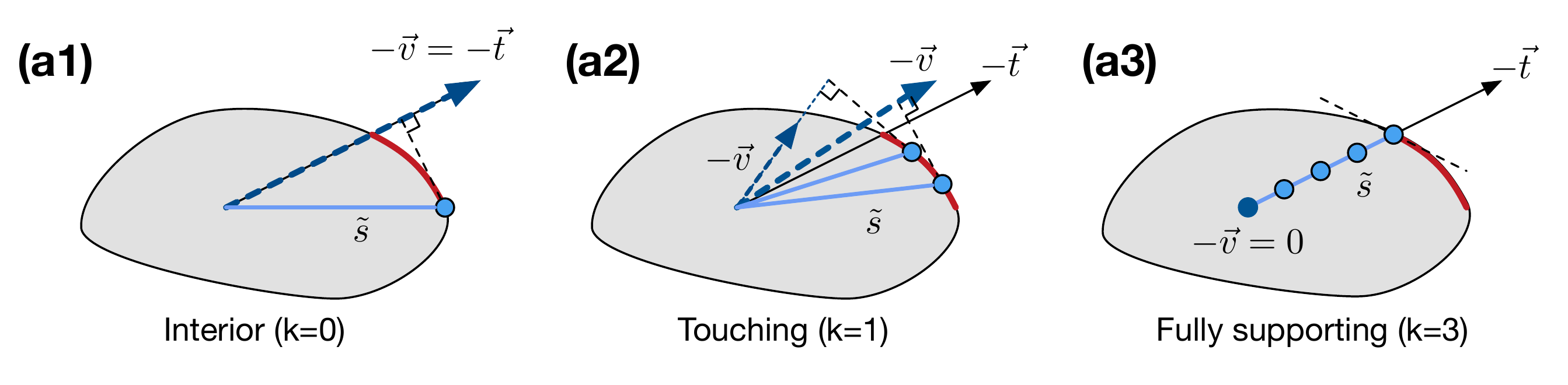} 
\par\end{centering}
\begin{centering}
\includegraphics[width=1\columnwidth]{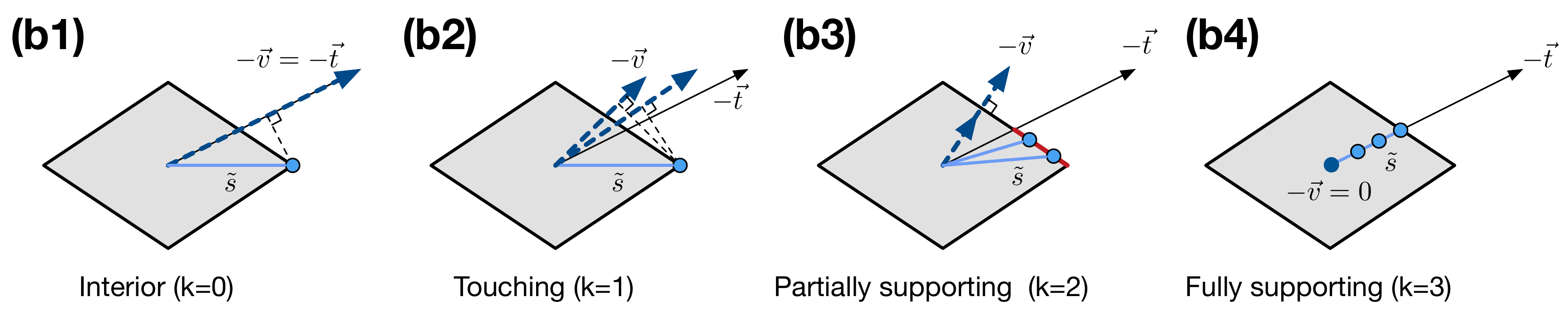} 
\par\end{centering}
\caption{Determining the anchor points of the Gaussian distributed vector $(\vec{t},t_{0})$
onto the convex hull of the manifold, denoted as $\tilde{s}(\vec{t},t_{0})$.
Here we show for the same vector $\vec{t}$, the change in $\tilde{s}(\vec{t},t_{0})$
as $t_{0}$ decreases from $+\infty$ to $-\infty$. (a) $D=2$ strictly
convex manifold. For sufficiently positive $t_{0}$ the vector $-\vec{t}$
obeys the constraint $g(\vec{t})+t_{0}>\kappa$, hence $-\vec{t}=-\vec{v}$
and the configuration corresponds to an interior manifold (support
dimension $k=0$). For intermediate values where ($t_{\text{touch}}>t_{0}-\kappa>t_{\text{fs}})$,
$(-\vec{t},-t_{0})$ violates the above constraints and $\tilde{s}$
is a point on the boundary of the manifold that maximizes the projection
on the manifold of a vector $(-\vec{v},-v_{0}$) that is the closest
to $(-\vec{t},-t_{0})$ and obeys $g(\vec{v})+t_{0}+\lambda=\kappa$.
Finally, for larger values of $-t_{0}$, $\vec{v}=0$ and $\tilde{s}$
is a point in the interior of the manifold in the direction of $-\vec{t}$
(fully supporting with $k=3$). (b) $D=2$ square manifold. Here,
in both the interior and touching regimes, $\tilde{s}$ is a vertex
of the square. In the fully supporting regime, the anchor point is
in the interior and collinear to $-\vec{t}.$ There is also a partially
supporting regime when $-t_{0}$ is slightly below $t_{\text{fs}}$.
In this regime, $-\vec{v}$ is perpendicular to one of the edges and
$\tilde{s}$ resides on an edge, corresponding to manifolds whose
intersection with the margin planes are edges $(k=2)$. \label{fig:Polar-Geometry}}
\end{figure}

\textbf{Fully supporting manifolds ($k=D+1$):} When $t_{0}$ is sufficiently
negative, $v_{0}=\kappa$, $\vec{v}=0$ and $\lambda=-t_{0}+\kappa$.
The anchor point $\tilde{s}(\vec{T})$ which obeys the KKT equations
resides in the interior of $\text{conv}\left(\mathcal{S}\right)$,
\begin{equation}
\tilde{s}(\vec{t},\,t_{0})=\frac{-\vec{t}}{\kappa-t_{0}}\label{eq:svsVT}
\end{equation}
For a fixed $\vec{t}$, $t_{0}$ must be negative enough, $t_{0}-\kappa<t_{\text{fs}}$,
where 
\begin{equation}
t_{\text{fs}}(\vec{t})=\arg\max\left\{ t_{0}\mid\frac{-\vec{t}}{\kappa-t_{0}}\in\text{conv\ensuremath{\left(\mathcal{S}\right)} }\right\} \label{eq:tembedGen}
\end{equation}
guaranteeing that $\tilde{s}(\vec{t},t_{0})\in\text{conv}\left(\mathcal{S}\right)$.
The contribution of this regime to the capacity is

\begin{equation}
F(\vec{T})=\left\Vert \vec{v}-\vec{t}\right\Vert ^{2}+(v_{0}-t_{0})^{2}=\left\Vert \vec{t}\right\Vert ^{2}+(\kappa-t_{0})^{2},\label{eq:F_full Supp}
\end{equation}

see Eq. \eqref{eq:kappaG}. Finally, for values of $t_{0}$ such that
$t_{\text{fs}}(\vec{t})\leq t_{0}-\kappa\leq t_{\text{touch}}(\vec{t})$,
the manifolds are partially supporting with support dimension $1\leq k\leq D$.
Examples of different supporting regimes are illustrated in Figure
\ref{fig:Polar-Geometry}.

\subsection{Effects of size and margin \label{subsec:Size-and-margin-1}}

We now discuss the effect of changing manifold size and the imposed
margin on both capacity and geometry. As described by Eq. \eqref{eq:rMmu},
change in the manifold size corresponds to scaling every $\tilde{s}$
by a scalar $r$.

\textbf{Small size:} If $r\rightarrow0,$ the manifold shrinks to
a point (the center), whose capacity reduces to that of isolated points.
However, in the case where $D\gg1$ , the capacity may be affected
by the manifold structure even if $r\ll1$, see section \ref{sec:High-Dimensional-Manifolds}.
Nevertheless, the underlying support structure is simple. When $r$
is small, the manifolds have only two support configurations. For
$t_{0}-\kappa>0,$ the manifold is interior with: $\lambda=0$ , $v_{0}=t_{0}$
and $\vec{v}=\vec{t}$. When $t_{0}-\kappa<0$, the manifold becomes
touching with support dimension $k=1$. In that case, because $\tilde{s}$
has a small magnitude, $\vec{v}\approx\vec{t}$, $\lambda\approx\kappa-t_{0}$,
and $v_{0}\approx\kappa$. Thus, in both cases, $\vec{v}$ is close
to the Gaussian vector $\vec{t}$. The probability of other configurations
vanishes.

\textbf{Large size:} In the large size limit $r\rightarrow\infty$,
separating the manifolds is equivalent to separating the affine subspaces.
As we show in Appendix \ref{sec:AppendixC_LargeLimit}, when $r\gg1$,
there are two main support structures. With probability $H\left(-\kappa\right)=\int_{-\kappa}^{\infty}Dt_{0}$
the manifolds are fully supporting, namely the underlying affine subspaces
are parallel to the margin plane. This regime contributes to the inverse
capacity an amount $H\left(-\kappa\right)D+\alpha_{0}^{-1}\left(\kappa\right)$.
The other regimes (touching or partially supporting) are such that
the angle between the affine subspace and the margin plane is \emph{almost
}zero, and contribute an amount $H(\kappa)D$ to the inverse capacity.
Combining the two contributions, we obtain for large sizes, $\alpha_{\text{M}}^{-1}=D+\alpha_{0}^{-1}(\kappa)$,
consistent with Eq. \eqref{eq:PNboundGardner}.

\textbf{Large margin:} For a fixed $\vec{T}$, Eq. \eqref{eq:tembedGen}
implies that larger $\kappa$ increases the probability of being in
the supporting regimes. Increasing $\kappa$ also shrinks the magnitude
of $\tilde{s}$ according to Eq. \eqref{eq:svsVT}. Hence, when $\kappa\gg1$,
the capacity becomes similar to that of $P$ random points and the
corresponding capacity is given by $\alpha_{\text{M}}\approx\kappa^{-2}$,
independent of manifold geometry.

\textbf{Manifold centers}: The theory of manifold classification described
in Sec. \ref{sec:Statistical-Mechanics} does not require the notion
of a manifold center. However, understanding how scaling the manifold
sizes by a parameter $r$ in \eqref{eq:rMmu} affects their capacity,
the center points about which the manifolds are scaled need to be
defined. For many geometries, the center is a point of symmetry such
as for an ellipsoid. For general manifolds, a natural definition would
be the center of mass of the anchor points $\tilde{S}(\vec{T})$ averaging
over the Gaussian measure of $\vec{T}$. We will adopt here a simpler
definition for the center provided by the \emph{Steiner point} for
convex bodies \citep{grunbaum1969convex}, $\vec{S}^{0}=(\vec{s}^{0},1)$
with 
\begin{equation}
\vec{s}^{0}=\left\langle \nabla g_{{\cal \mathcal{S}}}(\vec{t})\right\rangle _{\vec{t}}\label{eq:center}
\end{equation}
and the expectation is over the Gaussian measure of $\vec{t}\in\mathbb{R}^{D}$.
This definition coincides with the center of mass of the anchor points
when the manifold size is small. Furthermore, it is natural to define
the geometric properties of manifolds in terms of \emph{centered manifolds}
where the manifolds are shifted within their affine subspace so that
the center and orthogonal translation vector coincide, i.e. $\vec{s}^{0}=\vec{c}$
with $\left\Vert \vec{s}^{0}\right\Vert =\left\Vert \vec{c}\right\Vert =1$.
This means that all lengths are defined relative to the distance of
the centers from the origin and the \emph{$D$}-dimensional intrinsic
vectors $\vec{s}$ give the offset relative to the manifold center.

\subsection{Manifold anchor geometry \label{subsec:Non-gaussian-Geometry}}

The capacity equation \eqref{eq:amMgen} motivates defining geometrical
measures of the manifolds, which we call \textit{manifold anchor}
geometry. Manifold anchor geometry is based on the statistics of the
anchor points $\tilde{s}(\vec{t},t_{0})$ induced by the Gaussian
random vector $(\vec{t},t_{0})$ which are relevant to the capacity
in Eq. \eqref{eq:amMgen}. These statistics are sufficient for determining
the classification properties and supporting structures associated
with the maximum margin solution. We accordingly define the manifold
anchor radius and dimension as:

\textbf{Manifold anchor radius:} denoted $R_{\text{M}}$, is defined
by the mean squared length of $\tilde{s}(\vec{t},t_{0})$, 
\begin{equation}
R_{\text{M}}^{2}=\left\langle \left\Vert \tilde{s}(\vec{t},t_{0})\right\Vert ^{2}\right\rangle _{\vec{t,}t_{0}}\label{eq:RM}
\end{equation}
\textbf{Manifold anchor dimension:} $D_{\text{M}}$, is given by 
\begin{equation}
D_{\text{M}}=\left\langle \left(\vec{t}\cdot\hat{s}(\vec{t,}t_{0})\right)^{2}\right\rangle _{\vec{t,}t_{0}}\label{eq:DM}
\end{equation}
where $\hat{s}$ is a unit vector in the direction of $\tilde{s}$.
The anchor dimension measures the angular spread between $\vec{t}$
and the corresponding anchor point $\tilde{s}$ in $D$ dimensions.
Note that the manifold dimension obeys $D_{\text{M}}\leq\left\langle \left\Vert \vec{t}\right\Vert ^{2}\right\rangle =D$.
Whenever there is no ambiguity, we will call $R_{\text{M}}$ and $D_{\text{M}}$
the manifold radius and dimension, respectively.

These geometric descriptors offer a rich description of the manifold
properties relevant for classification. Since $\vec{v}$ and $\tilde{s}$
depend in general on $t_{0}-\kappa$, the above quantities are averaged
not only over $\vec{t}$ but also over $t_{0}$. For the same reason,
the manifold anchor geometry also depends upon the imposed margin.

\subsection{Gaussian geometry\label{subsec:Gaussian-Geometry}}

\begin{figure}
\begin{centering}
\includegraphics[width=1\columnwidth]{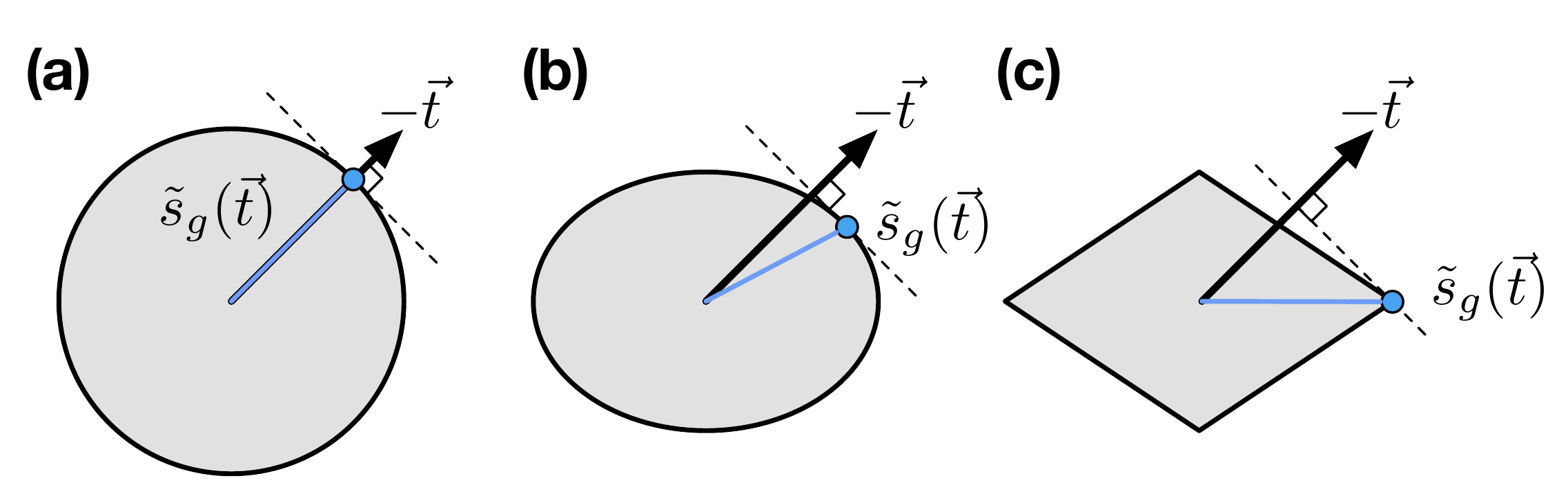} 
\par\end{centering}
\caption{Gaussian anchor points: Mapping from $\vec{t}$ to points $\tilde{s}_{g}(\vec{t})$,
showing the relation between $-\vec{t}$ and the point on the manifold
that touches a hyperplane orthogonal to $\vec{t}$ . $D=2$ manifolds
shown: (a) circle; (b) $\ell_{2}$ ellipsoid; (c) polytope manifold.
In (c), only for values of $\vec{t}$ of measure zero (when $\vec{t}$
is exactly perpendicular to an edge), $\tilde{s}_{g}(\vec{t})$ will
lie along the edge, otherwise it coincides with a vertex of the polytope.
In all cases, $\tilde{s}_{g}(\vec{t})$ will be in the interior of
the convex hulls only for $\vec{t}=0$. Otherwise, it is restricted
to their boundary.\label{fig:Gaussian-geometry}}
\end{figure}

We have seen that for small manifold sizes, $\vec{v}\approx\vec{t}$,
and the anchor points $\tilde{s}$ can be approximated by $\tilde{s}(\vec{t})=\nabla g_{S}(\hat{t})$.
Under these conditions, the geometry simplifies as shown in Fig. \ref{fig:Gaussian-geometry}.
For each Gaussian vector $\vec{t}\in\mathbb{R}^{D}$, $\tilde{s}(\vec{t})$
is the point on the manifold that first touches a hyperplane normal
to -$\vec{t}$ as it is translated from infinity towards the manifold.
Aside from a set of measure zero, this touching point is a unique
point on the boundary of $\mathcal{\text{conv(}S})$. This procedure
is similar to that used to define the well known \emph{Gaussian mean-width},
which in our notation equals $w(\mathcal{S})=-2\left\langle g_{\mathcal{S}}(\vec{t})\right\rangle _{\vec{t}}$\citep{giannopoulos2000convex}.
Note that for small sizes, the touching point does not depend on $t_{0}$
or $\kappa$ so its statistics are determined only by the shape of
$\text{conv}\left(\mathcal{S}\right)$ relative to its center. This
motivates defining a simpler manifold Gaussian geometry denoted by
the subscript $g$, which highlights its dependence over a $D$ dimensional
Gaussian measure for $\vec{t}$.

\textbf{Gaussian radius:} denoted by $R_{\text{g}}$, measures the
mean square amplitude of the Gaussian anchor point, $\tilde{s}_{g}(\vec{t})=\nabla g_{S}(\hat{t})$:
\begin{equation}
R_{\text{g}}^{2}=\left\langle \left\Vert \tilde{s}_{g}\left(\vec{t}\right)\right\Vert ^{2}\right\rangle _{\vec{t}}\label{eq:RMLowr-2}
\end{equation}

where the expectation is over $D$-dimensional Gaussian $\vec{t}$.

\textbf{Gaussian dimension:} $D_{\text{g}}$, is defined by, 
\begin{equation}
D_{\text{g}}=\left\langle \left(\vec{t}\cdot\hat{s}_{g}\left(\vec{t}\right)\right)^{2}\right\rangle _{\vec{t}}\label{eq:DGaussian}
\end{equation}
where $\hat{s}_{g}$ is the unit vector in the direction of $\tilde{s}_{g}$.
While $R_{\text{g}}^{2}$ measures the total variance of the manifold,
$D_{\text{g}}$ measures the angular spread of the manifold, through
the statistics of the angle between $\vec{t}$ and its Gaussian anchor
point. Note that $D_{\text{g}}\leq\left\langle \left\Vert \vec{t}\right\Vert ^{2}\right\rangle =D$.

It is important to note that even in this limit, our geometrical definitions
are not equivalent to conventional geometrical measures such as the
longest chord or second order statistics induced from a uniform measure
over the boundary of $\text{conv}\left(\mathcal{S}\right)$. For the
special case of $D$-dimensional $\ell_{2}$ balls with radius $R$,
$\tilde{s}_{g}\left(\vec{t}\right)$ is the point on the boundary
of the ball in the direction of $\vec{t}$ so that $R_{\text{g}}=R$
and $D_{\text{g}}=D$. However, for general manifolds, $D_{\text{g}}$
can be much smaller than the manifold affine dimension $D$ as illustrated
in some examples later.

We recapitulate the essential difference between the Gaussian geometry
and the full manifold anchor geometry. In the Gaussian case, the radius
is an intrinsic property of the shape of the manifold in its affine
subspace and is invariant to changing its distance from the origin.
Thus, scaling the manifold by a global scale factor $r$ as defined
in Eq. \eqref{eq:rMmu} results in scaling $R_{\text{g}}$ by the
factor $r$. Likewise, the dimensionality $D_{\text{g}}$ is invariant
to a global scaling of the manifold size. In contrast, the anchor
geometry does not obey such invariance for larger manifolds. The reason
is that the anchor point depends on the longitudinal degrees of freedom,
namely on the size of the manifold relative to the distance from the
center. Hence, $R_{\text{M}}$ need not scale linearly with $r$,
and $D_{\text{M}}$ will also depend on $r$. Thus, the anchor geometry
can be viewed as describing the general relationship between the signal
(center distance) to noise (manifold variability) in classification
capacity. We also note that the manifold anchor geometry automatically
accounts for the rich support structure described in Section \ref{subsec:Non-gaussian-Geometry}.
In particular, as $t_{0}$ decreases, the statistics of the anchor
points change from being concentrated on the boundary of $\text{conv\ensuremath{\left(\mathcal{S}\right)}}$
to its interior. Additionally for manifolds which are not strictly
convex and intermediate values of $t_{0}$, the anchor statistics
become concentrated on $k$-dimensional facets of the convex hull
corresponding to partially supported manifolds.

\begin{figure}
\begin{centering}
\includegraphics[width=1\columnwidth]{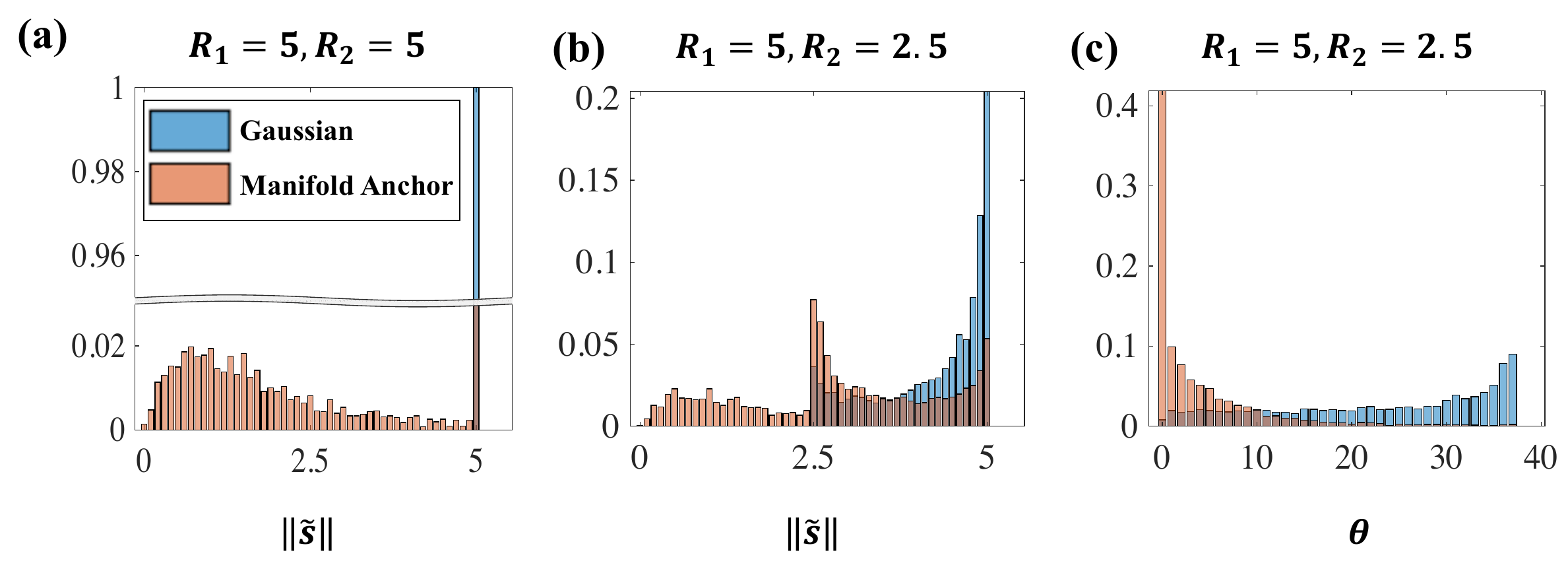} 
\par\end{centering}
\caption{\textcolor{black}{Distribution of $\left\Vert \tilde{s}\right\Vert $
(norm of manifold anchor vectors) and $\theta=\angle(-\vec{t},\tilde{s})$
for $D=2$ ellipsoids.}\textbf{\textcolor{black}{{} }}\textcolor{black}{(a)
Distribution of $\tilde{s}$ for $\ell_{2}$ ball with $R=5$. The
Gaussian geometry is peaked at $\left\Vert \tilde{s}\right\Vert =r$
(blue), while $\left\Vert \tilde{s}\right\Vert <r$ has non-zero probability
for the manifold anchor geometry (red). (b-c) $D=2$ ellipsoids, with
radii $R_{1}=5$ and $R_{2}=\frac{1}{2}R_{1}$. (b) Distribution of
$\left\Vert \tilde{s}\right\Vert $ for Gaussian geometry (blue) and
anchor geometry (red). (c) Corresponding distribution of $\theta=\angle(-\vec{t},\tilde{s})$.
\label{fig:DistS_2D_ELP} }}
\end{figure}

We illustrate the difference between the two geometries in Fig. \ref{fig:DistS_2D_ELP}
with two simple examples: a $D=2$ $\ell_{2}$ ball and a $D=2$ $\ell_{2}$
ellipse. In both cases we consider the distribution of $\left\Vert \tilde{s}\right\Vert $
and $\theta=\cos^{-1}\left(-\hat{t}\cdot\hat{s}\right)$, the angle
between $-\vec{t}$ and $\tilde{s}$. For the ball with radius $r$,
the vectors $-\vec{t}$ and $\tilde{s}$ are parallel so the angle
is always zero. For the manifold \emph{anchor} geometry, $\tilde{s}$
may lie \emph{inside} the ball in the fully supporting region. Thus
the distribution of $\left\Vert \tilde{s}\right\Vert $ consists of
a mixture of a delta function at $r$ corresponding to the interior
and touching regions and a smoothly varying distribution corresponding
to the fully supporting region.

Fig. \ref{fig:DistS_2D_ELP} also shows the corresponding distributions
for a two dimensional ellipsoid with major and minor radius, $R_{1}=r$
and $R_{2}=\frac{1}{2}r$. For the Gaussian geometry, the distribution
of $\left\Vert \tilde{s}\right\Vert $ has finite support between
$R_{1}$ and $R_{2}$, whereas the manifold anchor geometry has support
also below $R_{2}$. Since $\vec{t}$ and $\tilde{s}$ need not be
parallel, the distribution of the angle varies between zero and $\frac{\pi}{4}$,
with the manifold anchor geometry being more concentrated near zero
due to contributions from the fully supporting regime. In Section
\eqref{sec:Sparse-Labels}, we will show how the Gaussian geometry
becomes relevant even for larger manifolds when the labels are highly
imbalanced, i.e. sparse classification.

\subsection{Geometry and classification of high dimensional manifolds\label{sec:High-Dimensional-Manifolds}}

In general, linear classification as expressed in Eq. \eqref{eq:amMgen}
depends on high order statistics of the anchor vectors $\tilde{s}(\vec{t},t_{0}).$
Surprisingly, our analysis shows that for high dimensional manifolds,
the classification capacity can be described in terms of the statistics
of $R_{\text{M}}$ and $D_{\text{M}}$ alone. This is particularly
relevant as we expect that in many applications, the affine dimension
of the manifolds is large. More specifically, we define high-dimensional
manifolds as manifolds where the manifold dimension is large, i.e.,
$D_{\text{M}}\gg1$ (but still finite in the thermodynamic limit).
In practice we find that $D_{\text{M}}\gtrsim5$ is sufficient. Our
analysis below elucidates the interplay between size and dimension,
namely, how small $R_{\text{M}}$ needs to be for high dimensional
manifolds to have a substantial classification capacity.

In the high-dimensional regime, the mean field equations simplify
due to self averaging of terms involving sums of components $t_{i}$
and $\tilde{s}_{i}$. The two quantity, $\vec{t}\cdot\tilde{s}$,
that appears in the capacity, \eqref{eq:amMgen}, can be approximated
as $-\vec{t}\cdot\tilde{s}\approx\kappa_{\text{M}}$ where we introduce
the\emph{ effective manifold }margin\emph{ $\kappa_{\text{M}}=R_{\text{M}}\sqrt{D_{\text{M}}}$
. }Combined with $\left\Vert \tilde{s}\right\Vert ^{2}\approx R_{\text{M}}^{2}$
, we obtain

\begin{equation}
\alpha_{\text{M}}(\kappa)\approx\alpha_{0}\left(\frac{\kappa+\kappa_{\text{M}}}{\sqrt{1+R_{\text{M}}^{2}}}\right)\label{eq:capacityLargeD}
\end{equation}
where $\alpha_{0}$ is the capacity for $P$ random points, Eq. \eqref{eq:alpha0}.
To gain insight into this result we note that the effective margin
on the center is its mean distance from the point closest to the margin
plane, $\tilde{s}$, which is roughly the mean of $-\vec{t}\cdot\tilde{s}\approx R_{\text{M}}\sqrt{D_{\text{M}}}$.
The denominator in Eq. \eqref{eq:capacityLargeD} indicates that the
margin needs to be scaled by the input norm, $\sqrt{1+R_{\text{M}}^{2}}.$
In Appendix \ref{subsec:B.2.AppendixL2Balls}, we show that Eq. \eqref{eq:capacityLargeD}
can also be written as

\begin{equation}
\alpha_{\text{M}}(\kappa)\approx\alpha_{\text{Ball}}\left(\kappa,\,R_{\text{M}},\,D_{\text{M}}\right)\label{eq:capacityLargeD-1}
\end{equation}

namely, the classification capacity of a general high dimensional
manifold is well approximated by that of $\ell_{2}$ balls with dimension
$D_{\text{M}}$ and radius $R_{\text{M}}$.

\textbf{Scaling regime: }Eq. \eqref{eq:capacityLargeD} implies that
to obtain a finite capacity in the high-dimensional regime, the effective
margin $\kappa_{\text{M}}=R_{\text{M}}\sqrt{D_{\text{M}}}$ needs
to be of order unity which requires the radius to be small, scaling
for large $D_{\text{M}}$ as $R_{\text{M}}=O\left(D_{\text{M}}^{-\frac{1}{2}}\right)$.
In this\emph{ scaling regime}, the calculation of the capacity and
geometric properties are particularly simple. As argued above, when
the radius is small, the components of $\tilde{s}$ are small, hence
$\vec{v}\approx\vec{t}$, and the Gaussian statistics for the geometry
suffice. Thus, we can replace $R_{\text{M}}$ and $D_{\text{M}}$
in Eqs. \eqref{eq:capacityLargeD}- \eqref{eq:capacityLargeD-1} by
$R_{\text{g}}$ and $D_{\text{g}},$ respectively. Note that in the
scaling regime, the factor proportional to $1+R_{\text{g}}^{2}$ in
Eq. \eqref{eq:capacityLargeD} is the next order correction to the
overall capacity since $R_{\text{g}}$ is small. Notably, the margin
in this regime, $\kappa_{\text{g}}=R_{\text{g}}\sqrt{D_{\text{g}}}$,
is equal to half the \emph{Gaussian mean width} of convex bodies,
$\kappa_{\text{g}}\approx\frac{1}{2}w(\mathcal{S})$ \citep{vershynin2015estimation}.
As for the support structure, since the manifold size is small, the
only significant contributions arise from the interior ($k=0$) and
touching ($k=1$) supports.

\textbf{Beyond the scaling regime:} When $R_{\text{g}}$ is not small,
the anchor geometry, $R_{\text{M}}$ and $D_{\text{M}}$, cannot be
adequately described by the Gaussian statistics, $R_{\text{g}}$ and
$D_{\text{g}}$. In this case, the manifold margin $\kappa_{\text{M}}=R_{\text{M}}\sqrt{D_{\text{M}}}$
is large and Eq. \eqref{eq:capacityLargeD} reduces to :

\begin{equation}
\alpha_{\text{M}}\approx\frac{1+R_{\text{M}}^{-2}}{D_{\text{M}}}\label{eq:highD R1}
\end{equation}
where we have used $\alpha_{0}\left(\kappa\right)\approx\kappa^{-2}$
for large margins and assumed that $\kappa\ll\kappa_{\text{M}}$.
For strictly convex high dimensional manifolds with $R_{\text{M}}=O(1)$,
only the touching regime ($k=1$) contributes significantly to the
geometry and hence to the capacity. For manifolds that are not strictly
convex, partially supporting solutions with $k\ll D$ can also contribute
to the capacity.

Finally, when $R_{\text{M}}$ is large, fully supporting regimes with
$k\approx D$ contribute to the geometry, in which case, the manifold
anchor dimension approaches the affine dimension $D_{\text{M}}\rightarrow D$
and Eqs. \eqref{eq:capacityLargeD} and \eqref{eq:highD R1}reduce
to $\alpha_{\text{M}}\approx\frac{1}{D}$, as expected.

\section{Examples}

\subsection{Strictly convex manifolds: $\ell_{2}$ ellipsoids\label{sec:l2-Manifolds} }

The family of $\ell_{2}$ ellipsoids are examples of manifolds which
are \emph{strictly convex}. Strictly convex manifold have smooth boundaries
that do not contain corners, edges, or flats (see Appendix \ref{sec:AppendixB:StrictlyConvex}).
Thus, the description of the anchor geometry and support structures
is relatively simple. The reason is that the anchor vectors $\tilde{s}(\vec{t},t_{0})$
correspond to either interior $(k=0)$, touching $(k=1)$, or fully
supporting $(k=D+1),$ while partial support $(1<k<D+1)$ is not possible.
The $\ell_{2}$ ellipsoid geometry can be solved analytically; nevertheless,
because it has less symmetry than the sphere, it exhibits some salient
properties of manifold geometry such as nontrivial dimensionality
and non-uniform measure of the anchor points.

We assume the ellipsoids are centered relative to their symmetry centers
as described in Sec. \ref{sec:Manifold-Geometry}, and can be parameterized
by the set of points: $M^{\mu}=\mathbf{x}_{0}^{\mu}+\sum_{i=1}^{D}s_{i}\mathbf{u}_{i}^{\mu}$
where 
\begin{align}
\mathcal{S} & =\left\{ \vec{s}\mid\sum_{i=1}^{D}\left(\frac{s_{i}}{R_{i}}\right)^{2}\le1\right\} .\label{eq:constraint_ellipsoid}
\end{align}

The components of $\mathbf{u}_{i}^{\mu}$ and of the ellipsoid centers
$\mathbf{x_{0}}^{\mu}$ are i.i.d. Gaussian distributed with zero
mean and variance $\frac{1}{\sqrt{N}}$ so that they are orthonormal
in the large $N$ limit. The radii $R_{i}$ represent the principal
radii of the ellipsoids relative to the center. The anchor points
$\tilde{s}(\vec{t},t_{0})$ can be computed explicitly (details in
the Appendix \ref{sec:AppendixC_l2_ell}), corresponding to three
regimes.

Interior $(k=0)$, occurs when $t_{0}-\kappa\geq t_{\text{touch}}(\vec{t})$
where, 
\begin{equation}
t_{\text{touch}}(\vec{t})=\sqrt{\sum_{i=1}^{D}R_{i}^{2}t_{i}^{2}}\label{eq:ttouch}
\end{equation}

Here $\lambda=0$, resulting in zero contribution to the inverse capacity.

The touching regime $(k=1)$ holds when $t_{\text{touch}}(\vec{t})>t_{0}-\kappa>t_{\text{fs}}(\vec{t})$
and 
\begin{equation}
t_{\text{fs}}(\vec{t})=-\sqrt{\sum_{i=1}^{D}\left(\frac{t_{i}}{R_{i}}\right)^{2}}\label{eq:tembed}
\end{equation}

Finally, the fully supporting regime $(k=D+1)$ occurs when $t_{\text{fs}}(\vec{t})>t_{0}-\kappa$.
The full expression for the capacity for $\ell_{2}$ ellipsoids is
given in Appendix, Section \ref{sec:AppendixC_l2_ell}. Below, we
focus on the interesting cases of ellipsoids with $D\gg1$.

\textbf{High-dimensional ellipsoids:} It is instructive to apply the
general analysis of high dimensional manifolds to ellipsoids with
$D\gg1$. We will distinguish between different size regimes by assuming
that all the radii of the ellipsoid are scaled by a global factor
$r$. In the high dimensional regime, due to self-averaging, the boundaries
of the touching and fully supporting transitions can be approximated
by, 
\begin{align}
t_{\text{touch}} & =\sqrt{\sum_{i=1}^{D}R_{i}^{2}}\\
t_{\text{fs}} & =-\sqrt{\sum_{i=1}^{D}R_{i}^{-2}}\label{eq:tembed_ellipsoids}
\end{align}
both independent of $\vec{t}$. Then as long as $R_{i}\sqrt{D}$ are
not large (see below), $t_{\text{fs}}\rightarrow-\infty$, and the
probability of fully supporting vanishes. Discounting the fully supporting
regime, the anchor vectors $\tilde{s}(\vec{t},t_{0})$ are given by,

\begin{equation}
\tilde{s}_{i}=-\lambda_{i}t_{i}\label{eq:sLargeDE}
\end{equation}

\begin{equation}
\lambda_{i}=\frac{R_{i}^{2}}{Z(1+R_{i}^{2})}\label{eq:lambda_i_ellipsoid}
\end{equation}
where the normalization factor $Z^{2}=\sum_{i=1}^{D}\frac{R_{i}^{2}}{(1+R_{i}^{2})^{2}}$
(see Appendix \ref{sec:AppendixC_l2_ell}) . The capacity for high-dimensional
ellipsoids, is determined via Eq. \eqref{eq:capacityLargeD} with
manifold anchor radius, 
\begin{equation}
R_{\text{M}}^{2}=\sum_{i=1}^{D}\lambda_{i}^{2}\label{eq:RE_Rorder}
\end{equation}
and anchor dimension, 
\begin{equation}
D_{\text{M}}=\frac{\left(\sum_{i=1}^{D}\lambda_{i}\right)^{2}}{\sum_{i=1}^{D}\lambda_{i}^{2}}\label{eq:DE_Rorder}
\end{equation}
Note that $R_{\text{M}}/r$ and $D_{\text{M}}$ are not invariant
to scaling the ellipsoid by a global factor $r$, reflecting the role
of the fixed centers.

\textbf{The anchor covariance matrix}: We can also compute the covariance
matrix of the anchor points. This matrix is diagonal in the principal
directions of the ellipsoid, and its eigenvalues are $\lambda_{i}^{2}$.
It is interesting to compare $D_{\text{M}}$ with a well known measure
of an effective dimension of a covariance matrix, \emph{the participation
ratio} \citep{rajan2010stimulus,litwin2017optimal}. Given a spectrum
of eigenvalues of the covariance matrix, in our notation $\lambda_{i}^{2}$,
we can define a \emph{generalized} participation ratio as $PR_{q}=\frac{\left(\sum_{i=1}^{D}\lambda_{i}^{q}\right)^{2}}{\sum_{i=1}^{D}\lambda_{i}^{2q}}$,
$q>0$. The conventional participation ratio uses $q=2$, whereas
$D_{\text{M}}$ uses $q=1$ .

\textbf{Scaling regime: }In the scaling regime where the radii are
small, the radius and dimension are equivalent to the Gaussian geometry:
\begin{equation}
R_{\text{g}}^{2}=\frac{\sum_{i=1}^{D}R_{i}^{4}}{\sum_{i=1}^{D}R_{i}^{2}}\label{eq:RE_Rorder1-1-1}
\end{equation}

\begin{equation}
D_{\text{g}}=\frac{\left(\sum_{i=1}^{D}R_{i}^{2}\right)^{2}}{\sum_{i=1}^{D}R_{i}^{4}}\label{eq:DE_Rorder1-1-1}
\end{equation}
and the effective margin is given as, 
\begin{equation}
\kappa_{\text{g}}=\left(\sum_{i=1}^{D}R_{i}^{2}\right)^{1/2}\label{eq:kappaElargeD-1}
\end{equation}

As can be seen, $D_{\text{g}}$ and $\frac{R_{\text{g}}}{r}$ are
invariant to scaling all the radii by $r$, as expected. It is interesting
to compare the above Gaussian geometric parameters with the statistics
induced by a uniform measure on the surface of the ellipsoid. In this
case, the covariance matrix has eigenvalues $\lambda_{i}^{2}=R_{i}^{2}$
and the total variance is equal to $\sum_{i}R_{i}^{2}$. In contrast,
in the Gaussian geometry, the eigenvalues of the covariance matrix,
$\lambda_{i}^{2}$, are proportional to $R_{i}^{4}$ . This and the
corresponding expression \eqref{eq:RE_Rorder1-1-1} for the squared
radius is the result of a non-uniform induced measure on the surface
of the ellipse in the anchor geometry, even in its Gaussian limit.

\textbf{Beyond the scaling regime: }When\textbf{ $R_{i}=O(1)$}, the
high dimensional ellipsoids all become touching manifolds since $t_{\text{touch}}\rightarrow-\infty$
and $t_{\text{fs}}\rightarrow\infty$. The capacity is small because
$\kappa_{\text{M}}\gg1$, and is given by Eq. \eqref{eq:highD R1}
with Eqs. \eqref{eq:RE_Rorder}-\eqref{eq:DE_Rorder}. Finally, when
all $R_{i}\gg1$, we have, 
\begin{equation}
R_{\text{M}}^{2}=\frac{D}{\sum_{i=1}^{D}R_{i}^{-2}}=\frac{1}{\langle R_{i}^{-2}\rangle_{i}}\label{eq:RE_Rorder1-1-1-1}
\end{equation}
Here $R_{\text{M}}$ scales with the global scaling factor, $r,$
$D_{\text{M}}=D$, and $\alpha^{-1}=D$. Although the manifolds are
touching, their angle relative to the margin plane is near zero. When
$R_{i}\approx\sqrt{D}$ or larger, the fully supporting transition
$t_{\text{fs}}$ becomes order one and the probability of fully supporting
is significant.

\begin{figure}
\begin{centering}
\includegraphics[width=1\columnwidth]{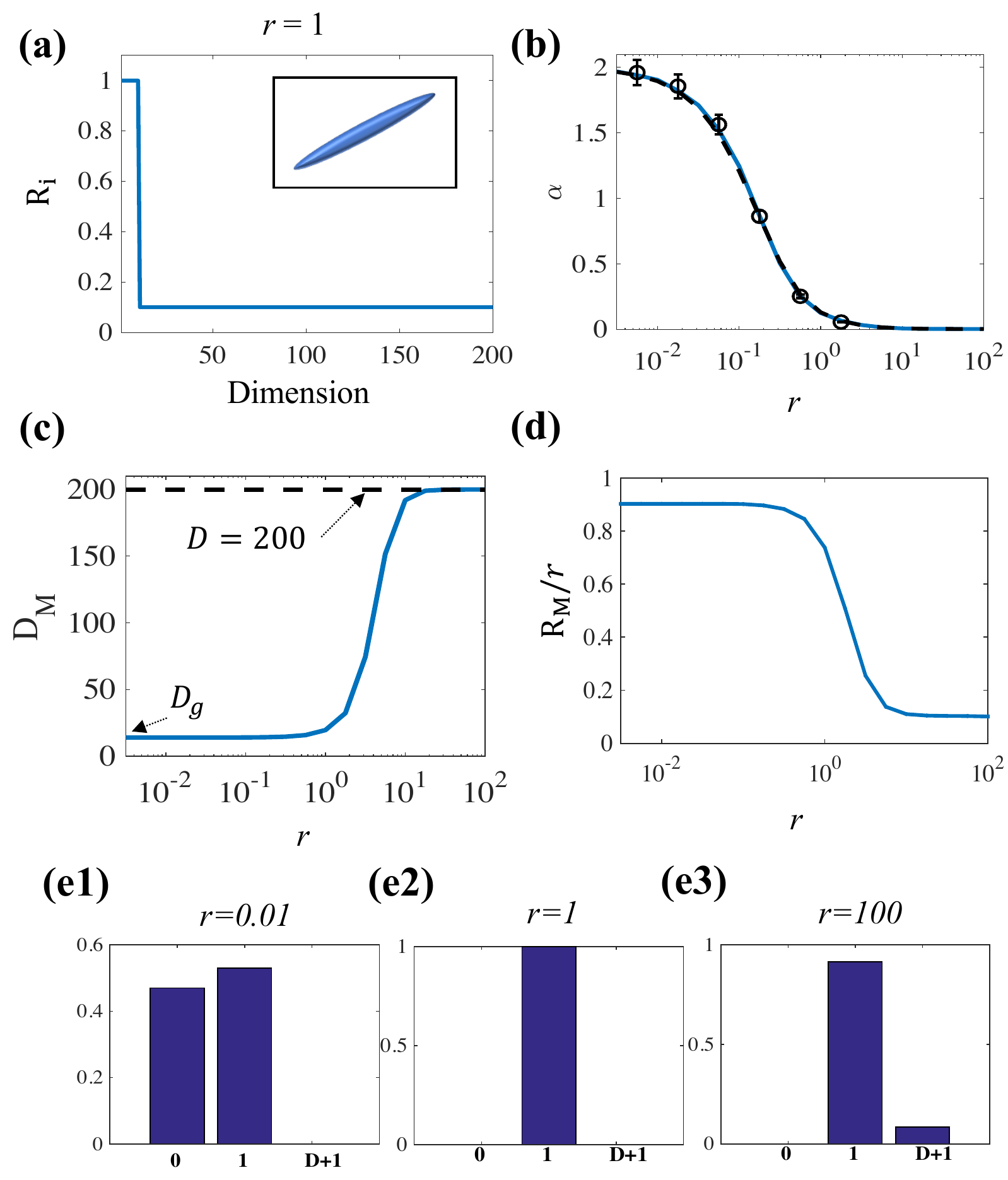} 
\par\end{centering}
\caption{Bimodal $\ell_{2}$ ellipsoids. (a) Ellipsoidal radii $R_{i}$ with
$r=1$ (b) Classification capacity as a\textcolor{black}{{} function
of scaling factor $r$: (blue lines) full mean field theory capacity,
(black dashed) approximation of the capacity given by equivalent ball
with $R_{\text{M}}$ and $D_{\text{M}}$, (circles) simulation capacity,
averaged over 5 repetitions, measured with 50 dichotomies per each
repetition. (c) Manifold dimension $D_{\text{M}}$ as a function of
$r$. (d) Manifold radius $R_{\text{M}}$ relative to $r$. (e) Fraction
of manifolds with support dimension $k$ for different values of $r$:
$k=0$ (interior), $k=1$ (touching), $k=D+1$ (fully supporting).
(e1)} Small $r=0.01,$ where most manifolds are interior or touching
(e2) $r=1$, where most manifolds are in the touching regime (e3)
$r=100$, where the fraction of fully supporting manifolds is 0.085,
predicted by $H(-t_{\text{fs}})=H(\sqrt{\sum_{i}R_{i}^{-2}})$ (Eq.
\ref{eq:tembed_ellipsoids}). \label{fig:Bimodal-Ellipsoids}}
\end{figure}

In Fig. \ref{fig:Bimodal-Ellipsoids} we illustrate the behavior of
high $D$-dimensional ellipsoids, using ellipsoids with a bimodal
distribution of their principal radii, $R_{i}$: $R_{i}=r,$ for $1\leq i\leq10$
and $R_{i}=0.1r,$ for $11\leq i\leq200$ (Fig. \ref{fig:Bimodal-Ellipsoids}(a)).
Their properties are shown as a function of the overall scale $r$.
Fig. \ref{fig:Bimodal-Ellipsoids}(b) shows numerical simulations
of the capacity, the full mean field solution as well as the spherical
high dimensional approximation (with $R_{\text{M}}$ and $D_{\text{M}}$).
These calculations are all in good agreement, showing the accuracy
of the mean field theory and spherical approximation. As seen in (b)-(d),
the system is in the scaling regime for $r<0.3$. In this regime,
the manifold dimension is constant and equals $D_{\text{g}}\approx14$,
as predicted by the participation ratio, Eq. \eqref{eq:DE_Rorder1-1-1},
and the manifold radius $R_{\text{g}}$ is linear with $r$, as expected
from Eq. \eqref{eq:RE_Rorder1-1-1}. The ratio $\frac{R_{\text{g}}}{r}\approx0.9$
is close to unity, indicating that in the scaling regime, the system
is dominated by the largest radii. For $r>0.3$ the effective margin
is larger than unity and the system becomes increasingly affected
by the full affine dimensionality of the ellipsoid, as seen by the
marked increase in dimension as well as a corresponding decrease in
$\frac{R_{\text{M}}}{r}$. For larger $r$, $D_{\text{M}}$ approaches
$D$ and $\alpha_{\text{M}}^{-1}=D$. Fig. \ref{fig:Bimodal-Ellipsoids}(e1)-(e3)
shows the distributions of the support dimension $0\le k\le D+1$.
In the scaling regime, the interior and touching regimes each have
probability close to $\frac{1}{2}$ and the fully supporting regime
is negligible. As $r$ increases beyond the scaling regime, the interior
probability decreases and the solution is almost exclusively in the
touching regime. For very high values of $r,$ the fully supporting
solution gains a substantial probability. Note that the capacity decreases
to approximately $\frac{1}{D}$ for a value of $r$ below where a
substantial fraction of solutions are fully supporting. In this case,
the touching ellipsoids all have very small angle with the margin
plane.

\begin{figure}
\begin{centering}
\includegraphics[width=1\columnwidth]{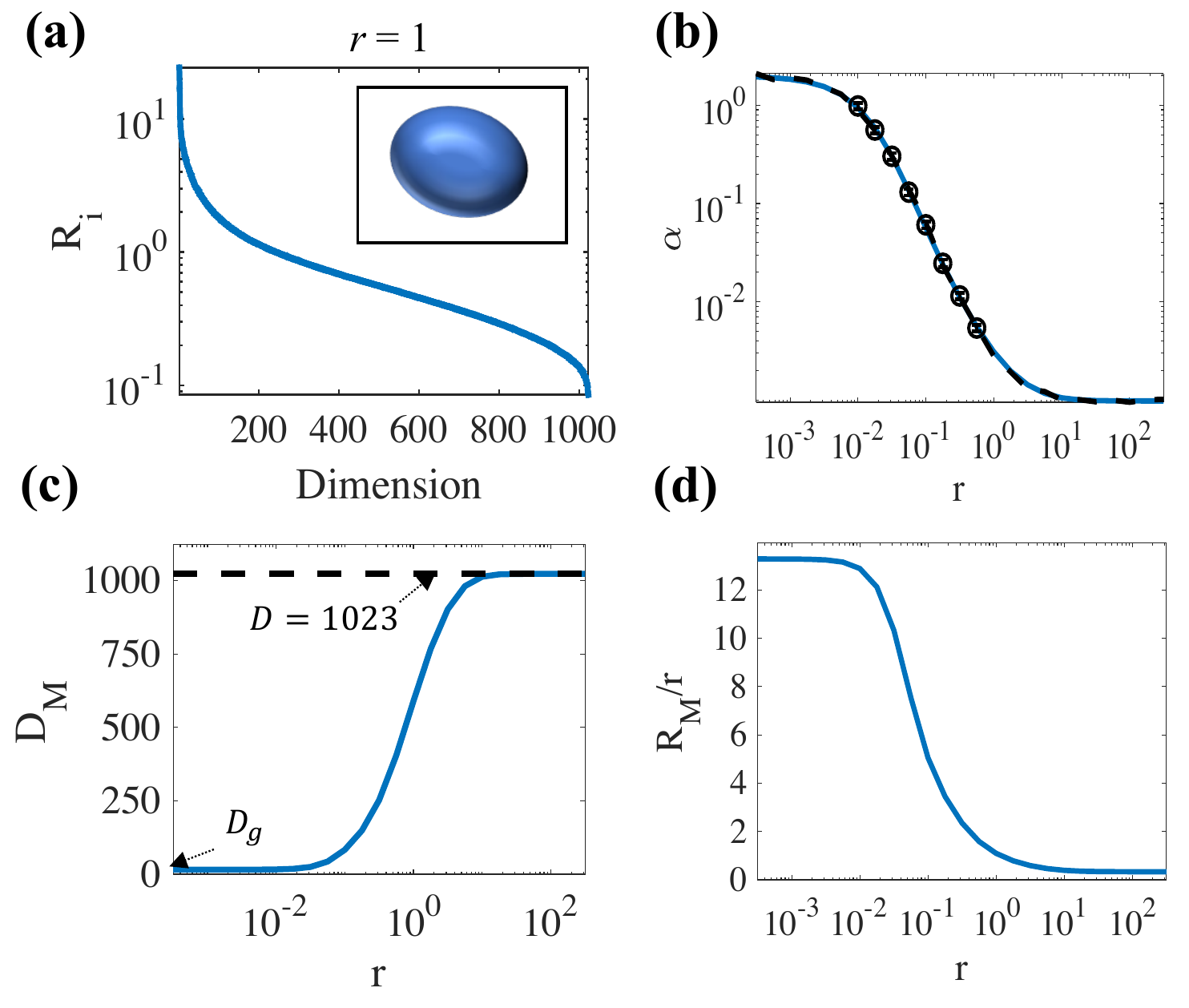} 
\par\end{centering}
\caption{Ellipsoids with radii computed from realistic image data. (a) SVD
spectrum, $R_{i}$, taken from the readout layer of GoogLeNet from
a class of ImageNet images with $N=1024$ and $D=1023.$ The radii
are scaled by a factor $r$: $R'_{i}=rR_{i}$ for (b) Classification
capacity as a function of $r$. \textcolor{black}{(blue lines) full
mean field theory capacity, (black dashed) approximation of the capacity
as that of a ball with $R_{\text{M}}$ and $D_{\text{M}}$ from the
theory for ellipsoids, (circles) simulation capacity, averaged over
5 repetitions, measured with 50 random dichotomies per each repetition.}
(c) Manifold dimension as a function of $r$. (d) Manifold radius
relative to the scaling factor $r,$ $R_{\text{M}}/r$ as a function
of $r$.\label{fig:Ellipsoids-with-Realistic}}
\end{figure}

Until now, we have assumed that the manifold affine subspace dimension
$D$ is finite in the limit of large ambient dimension, $N$. In realistic
data, it is likely that the data manifolds are technically full rank,
i.e. $D+1=N$, raising the question whether our mean field theory
is still valid in such cases. We investigate this scenario by computing
the capacity on ellipsoids containing a realistic distribution of
radii. We have taken as examples, a class of images from the ImageNet
dataset \citep{deng2009imagenet}, and analyzed the SVD spectrum of
the representations of these images in the last layer of a deep convolutional
network, GoogLeNet \citep{szegedy2015going}. The computed radii are
shown in Fig. \ref{fig:Ellipsoids-with-Realistic}(a) and yield a
value $D=N-1=1023$. In order to explore the properties of such manifolds,
we have scaled all the radii by an overall factor $r$ in our analysis.
Because of the decay in the distribution of radii, the Gaussian dimension
for the ellipsoid is only about $D_{\text{g}}\approx15$, much smaller
than $D$ or $N$, implying that for small $r$ the manifolds are
effectively low dimensional and the geometry is dominated by a small
number of radii. As $r$ increases above $r\gtrsim0.03$, $\kappa_{\text{M}}$
becomes larger than $1$ and the solution leaves the scaling regime,
resulting in a rapid increase in $D_{\text{M}}$ and a rapid falloff
in capacity as shown in Fig. \ref{fig:Ellipsoids-with-Realistic}(b-c).
Finally, for $r>10$, we have $\alpha_{\text{M}}^{-1}\approx D_{\text{M}}\approx D$
approaching the lower bound for capacity as expected. The agreement
between the numerical simulations and the mean field estimates of
the capacity illustrates the relevance of the theory for realistic
data manifolds that can be full rank.

\subsection{Convex polytopes: $\ell_{1}$ ellipsoids\label{sec:l1-Balls}}

The family of $\ell_{2}$ ellipsoids represent manifolds that are
smooth and strictly convex. On the other hand, there are other types
of manifolds whose convex hulls are not strictly convex. In this section,
we consider a $D$-dimensional $\ell_{1}$ ellipsoids. They are prototypical
of \emph{convex polytopes} formed by the convex hulls of finite numbers
of points in $\mathbb{R}^{N}$. The $D$-dimensional $\ell_{1}$ ellipsoid,
is parameterized by radii $\left\{ R_{i}\right\} $ and specified
by the convex set $M^{\mu}=\mathbf{x}_{0}^{\mu}+\sum_{i=1}^{D}s_{i}\mathbf{u}_{i}^{\mu}$
, with: 
\begin{equation}
\mathcal{S}=\left\{ \vec{s}\mid\sum_{i=1}^{D}\frac{\left|s^{i}\right|}{R_{i}}\le1\right\} .
\end{equation}

Each manifold $M^{\mu}\in\mathbb{R}^{N}$ is centered at $\mathbf{x}_{0}^{\mu}$
and consists of a convex polytope with a finite number ($2D$) of
vertices: $\left\{ \mathbf{x}_{0}^{\mu}\pm R_{k}\mathbf{u}_{k}^{\mu},\,k=1,...,D\right\} $.
The vectors $\mathbf{u}_{i}^{\mu}$ specify the principal axes of
the $\ell_{1}$ ellipsoids. For simplicity, we consider the case of
$\ell_{1}$ balls when all the radii are equal: $R_{i}=R$. We will
concentrate on the cases when $\ell_{1}$ balls are high-dimensional;
the case for $\ell_{1}$ balls with $D=2$ was briefly described in
\citep{chung2016linear}. The analytical expression of the capacity
is complex due to the presence of contributions from all types of
supports, $1\leq k\leq D+1$. We address important aspects of the
high dimensional solution below.

\textbf{High-dimensional $\ell_{1}$ balls, scaling regime: }In the
scaling regime, we have $\vec{v}\approx\vec{t}$. In this case, we
can write the solution for the subgradient as: 
\begin{equation}
\tilde{s}_{i}(\vec{t})=\left\{ \begin{array}{cc}
-R\,\text{sign}\left(t_{i}\right), & \left|t_{i}\right|>\left|t_{j}\right|\forall j\ne i\\
0 & \text{otherwise}
\end{array}\right.
\end{equation}
In other words, $\tilde{s}(\vec{t})$ is a vertex of the polytope
corresponding to the component of $\vec{t}$ with the largest magnitude,
see Fig. \ref{fig:Gaussian-geometry}(c). The components of $\vec{t}$
are i.i.d. Gaussian random variables, and for large $D$, its maximum
component, $t_{max},$ is concentrated around $\sqrt{2\log D}$. Hence,
$D_{\text{g}}=\langle(\hat{s}\cdot\vec{t})^{2}\rangle=\langle t_{max}^{2}\rangle=2\log D$
which is much smaller than $D$. This result is consistent with the
fact that the Gaussian mean width of a $D$-dimensional $\ell_{1}$
ball scales with $\sqrt{\log D}$ and not with $D$ \citep{vershynin2015estimation}.
Since all the points have norm $R$, we have $R_{\text{g}}=R$, and
the effective margin is then given by $\kappa_{\text{g}}=R\sqrt{2\log D}$
, which is order unity in the scaling regime. In this regime, the
capacity is given by simple relation $\alpha_{\text{M}}=\alpha_{0}\left(\left[\kappa+R\sqrt{2\log D}\right]/\sqrt{1+R^{2}}\right)$.

\textbf{High-dimensional $\ell_{1}$ balls, $R=O(1)$: }When the radius
$R$ is small as in the scaling regime, the only contributing solution
is the \emph{touching} solution $(k=1)$. When $R$ increases, solutions
with all values of $k$, $1\leq k\leq D+1$ occur, and the support
can be any face of the convex polytope with dimension $k$. As $R$
increases, the probability distribution $p(k)$ over $k$ of the solution
shifts to larger values. Finally, for large $R,$ only two regimes
dominate: fully supporting ($k=D+1$) with probability $H\left(-\kappa\right)$
and partially supporting with $k=D$ with probability $H(\kappa$). 
\begin{center}
\begin{figure}
\begin{centering}
\includegraphics[width=1\columnwidth]{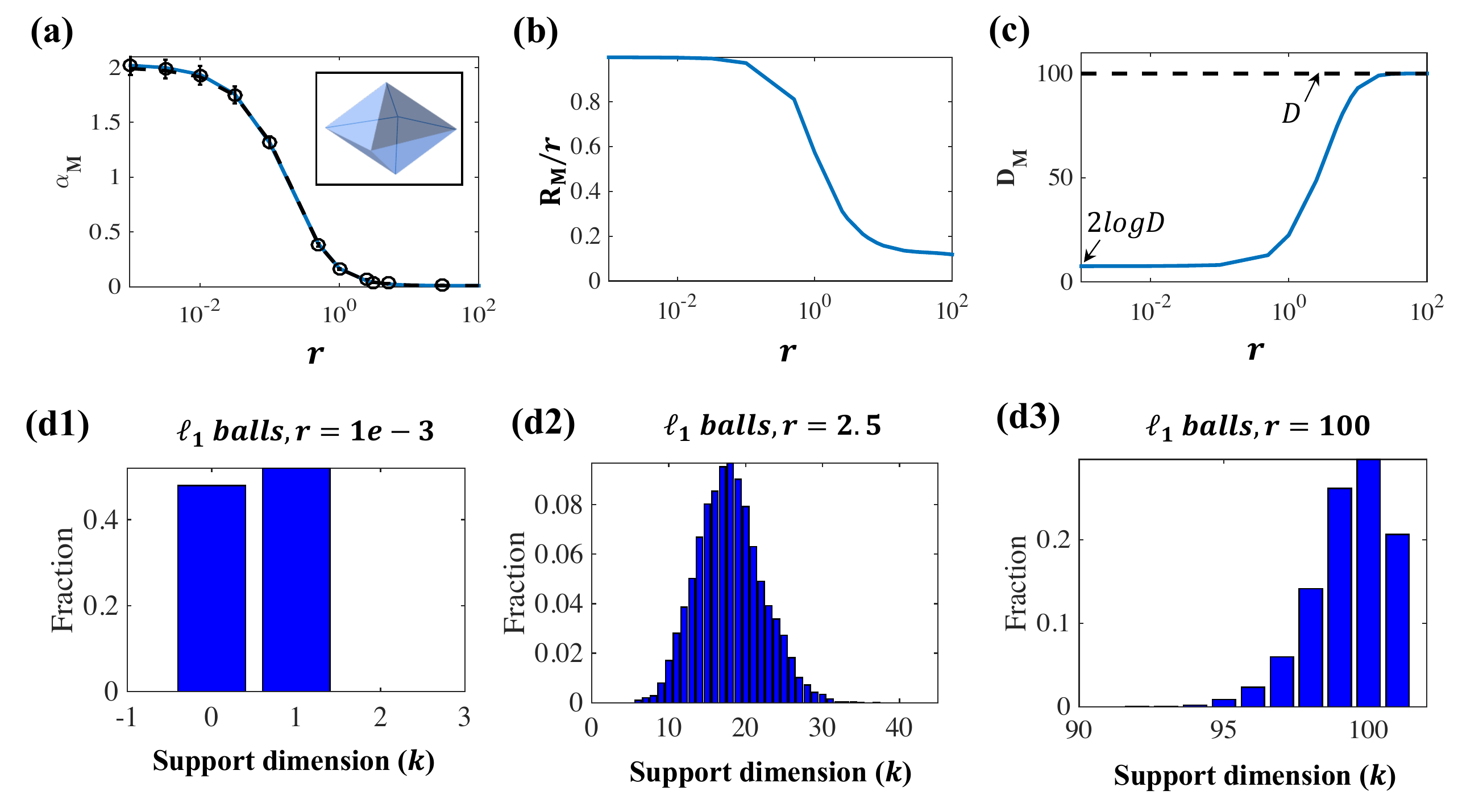} 
\par\end{centering}
\caption{Separability of $\ell_{1}$ balls. (a) Linear class\textcolor{black}{ification
capacity of $\ell_{1}$ balls as function of radius $r$ with $D=100$
and $N=200$: (blue) MFT solution, (black dashed) spherical approximation,
(circle) full numerical simulations. }(Inset) Illustration of $\ell_{1}$
\textcolor{black}{ball. (b) Manifold radius $R_{\text{M}}$ relative
to the actual radius $r$. (c) Manifold dimension $D_{\text{M}}$
as a function of $r$. In small $r$ limit, $D_{\text{M}}$ is approximately
$2\log D$ , while in large $r$, $D_{\text{M}}$ is close to $D$,
showing how the solution is orthogonal to the manifolds when their
sizes are large. (d1-d3) Distribution of support dimensions: (d1)
$r=0.001$, where most manifolds are either interior or touching,
(d2) $r=2.5$, where the support dimension has a peaked distribution,
(d3) $r=100$, where most manifolds are close to being fully supporting.
\label{fig:L1-balls}}}
\end{figure}
\par\end{center}

We illustrate the behavior of $\ell_{1}$ balls with radius $r$ and
affine dimension $D=100$. In Fig. \ref{fig:L1-balls}, (a) shows
the linear classification capacity as a function of $r$. When $r\rightarrow0$,
the manifold approaches the capacity for isolated points, $\alpha_{\text{M}}\approx2$,
and when $r\rightarrow\infty$, $\alpha_{\text{M}}\approx\frac{1}{D}=0.01$.
The numerical simulations demonstrate that despite the different geometry,
the capacity of the $\ell_{1}$ polytope is similar to that of a $\ell_{2}$
ball with radius $R_{\text{M}}$ and dimension $D_{\text{M}}$. In
(b), for the scaling regime when $r<0.1=\frac{1}{\sqrt{D}},$ we see
$R_{\text{M}}\approx r$, but when $r\gg1$, $R_{\text{M}}$ is much
smaller than $r$, despite the fact that all $R_{i}$ of the polytope
are equal. This is because when $r$ is not small, the various faces
and eventually interior of the polytope contribute to the anchor geometry.
In (c), we see $D_{\text{M}}\approx2\log D$ in the scaling regime,
while $D_{\text{M}}\rightarrow D$ as $r\rightarrow\infty$. In terms
of the support structures, when $r=0.001$ in the scaling regime (d1),
most manifolds are either interior or touching. For intermediate sizes
(d2), the support dimension is peaked at an intermediate value, and
finally for very large manifolds (d3), most polytope manifolds are
nearly fully supporting.

\subsection{Smooth nonconvex manifolds: Ring manifolds \label{sec:Orientation-Manifolds}}

Many neuroscience experiments measure the responses of neuronal populations
to a continuously varying stimulus, with one or a small number of
degrees of freedom. A prototypical example is the response of neurons
in visual cortical areas to the orientation (or direction of movement)
of an object. When a population of neurons respond both to an object
identity as well as to a continuous physical variation, the result
is a set of smooth manifolds each parameterized by a single variable,
denoted $\theta$, describing continuous curves in $\mathbb{R}^{N}$.
Since in general the neural responses are not linear in $\theta$,
the curve spans more than one linear dimension. This smooth curve
is not convex, and is endowed with a complex non-smooth convex hull.
It is thus interesting to consider the consequences of our theory
on the separability of smooth but non-convex curves.

The simplest example, considered here is the case where $\theta$
corresponds to a periodic angular variable such as orientation of
an image, and we call the resulting non-convex curve a \emph{ring}
manifold. We model the neuronal responses as smooth periodic functions
of $\theta$, which can be parameterized by decomposing the neuronal
responses in Fourier modes. Here, for each object, $M^{\mu}=\mathbf{x}_{0}^{\mu}+\sum_{i=1}^{D}s_{i}\mathbf{u}_{i}^{\mu}$
where $\mathbf{x}_{0}^{\mu}$ represents the mean (over $\theta$)
of the population response to the object. The different $D$ components
correspond to the different Fourier components of the angular response,
so that 
\begin{align}
s^{2n}(\theta) & =\frac{R_{n}}{\sqrt{2}}\cos(n\theta)\label{eq:s2n}\\
s^{2n-1}(\theta) & =\frac{R_{n}}{\sqrt{2}}\text{sin}(n\theta)\nonumber 
\end{align}
where $R_{n}$ is the magnitude of the $n$-th Fourier component for
$1\le n\le\frac{D}{2}$. The neural responses in Eq. \eqref{eq:d+1manifolds}
are determined by projecting onto the basis: 
\begin{align}
\mathbf{u}_{2n}^{\mu i} & =\sqrt{2}\cos(n\theta^{\mu i})\label{eq:u2n}\\
\mathbf{u}_{2n-1}^{\mu i} & =\sqrt{2}\sin(n\theta^{\mu i})\nonumber 
\end{align}
The parameters $\theta^{\mu i}$ are the preferred orientation angles
for the corresponding neurons and are assumed to be evenly distributed
between $-\pi\le\theta_{i}^{\mu}\le\pi$ (for simplicity, we assume
that the orientation tuning of the neurons are all of the same shape
and are symmetric around the preferred angle). The statistical assumptions
of our analysis assume that the different manifolds are randomly positioned
and oriented with respect to the others. For the ring manifold model,
this implies that the mean responses $\mathbf{x}_{0}^{\mu}$ are independent
random Gaussian vectors and also that the preferred orientation $\theta^{\mu i}$
angles are uncorrelated. With this definition, all the vectors $\vec{s}\in\mathcal{S}$
obey the normalization $\left\Vert \vec{s}\right\Vert =r$ where $r^{2}=\sum_{n=1}^{\frac{D}{2}}R_{n}^{2}$.
Thus, for each object the ring manifold is a closed non-intersecting
smooth curve residing on the surface of a $D$-dimensional sphere
with radius $r$ .

\begin{figure}
\begin{centering}
\includegraphics[width=1\columnwidth]{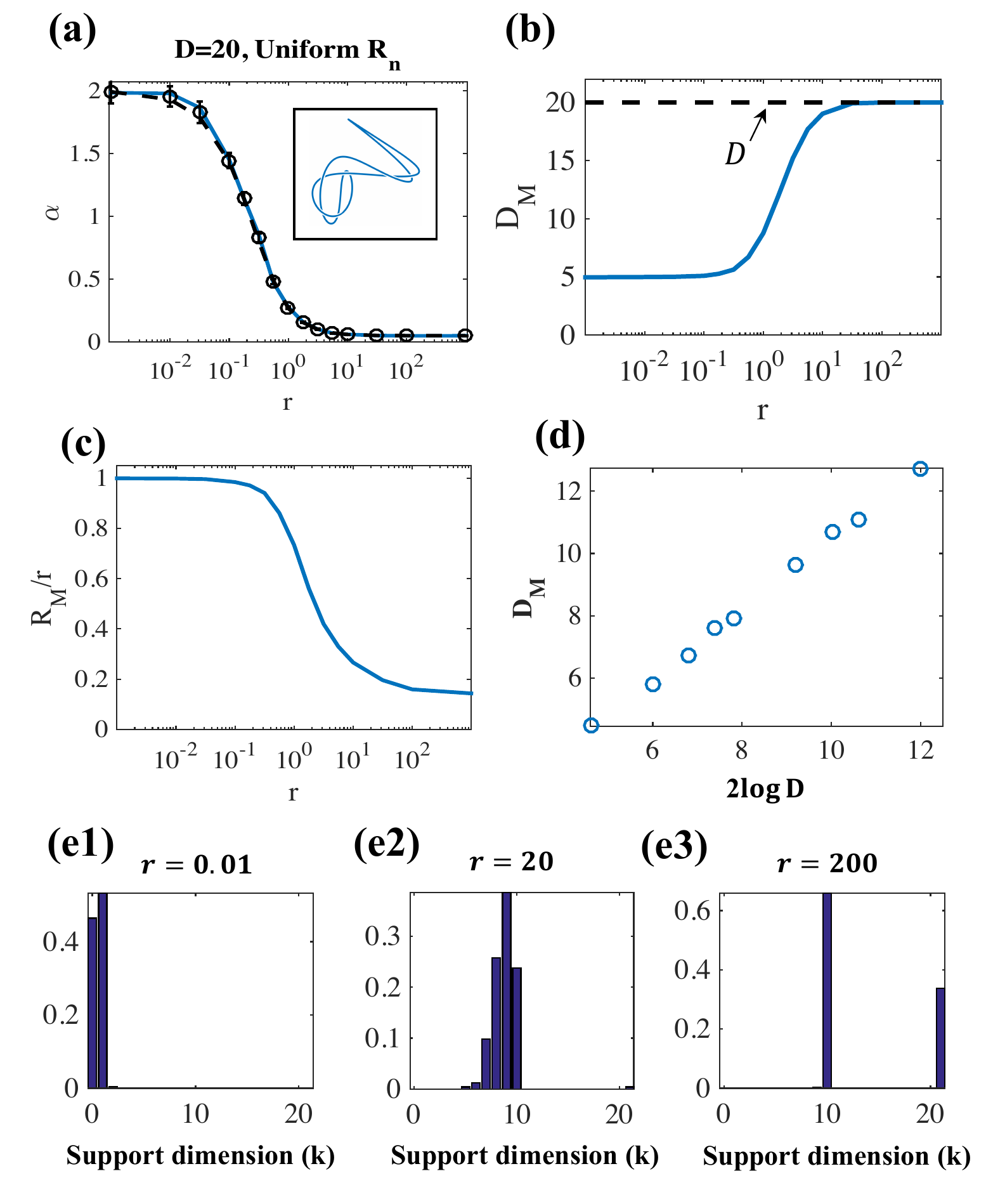} 
\par\end{centering}
\caption{Linear classification of $D$-dimensional ring manifolds, with uniform
$R_{n}=\sqrt{\frac{2}{D}}r$. (a) Classification capacity for $D=20$
as function of $r$ with $m=200$ test samples (details of numerical
simulations in SM, Sec. S9): (blue) mean field theory, (black dashed)
spherical approximation, (black circles) numerical simulations. (Inset)
Illustration of a ring manifold. (b) Manifold dimension $D_{\text{M}}$,
which shows $D_{\text{M}}\sim D$ in the large $r$ limit, showing
the orthogonality of the solution. (c) Manifold radius relative to
the scaling factor $r,$ $\frac{R_{\text{M}}}{r}$, as a function
of $r$. Fact that $\frac{R_{\text{M}}}{r}$ becomes small implies
that the manifolds are ``fully supporting'' to the hyperplane, showing
the small radius structure. (d) Manifold dimension grows with affine
dimension, $D$ , as $2\log D$ for small $r=\frac{1}{2\sqrt{2\log D}}$
in the scaling regime.(e1-e3) Distribution of support dimensions.
\textcolor{black}{(d1) $r=0.01$, where most manifolds are either
interior or touching, (d2) $r=20$, where the support dimension has
a peaked distribution, truncated at $\frac{D}{2}$ (d3) $r=200$,
where most support dimensions are $\frac{D}{2}$ or fully supporting
at $D+1$.}\label{fig:Ori_Uniform} }
\end{figure}

For the simplest case with $D=2$, the ring manifold is equivalent
to a circle in two dimensions. However, for larger $D$, the manifold
is not convex and its convex hull is composed of faces with varying
dimensions. In Fig. \ref{fig:Ori_Uniform}, we investigate the geometrical
properties of these manifolds relevant for classification as a function
of the overall scale factor $r$, where for simplicity we have chosen
$R_{n}=r$ for all $n$. The most striking feature is the small dimension
in the scaling regime, scaling roughly as $D_{\text{g}}\approx2\log D$.
This logarithmic dependence is similar to that of the $\ell_{1}$
ball polytopes. Then, as $r$ increases, $D_{\text{M}}$ increases
dramatically from $2\log D$ to $D$.

The similarity of the ring manifold to a convex polytope is also seen
in the support dimension $k$ of the manifolds. Support faces of dimension
$k\neq0,1,D+1$ are seen, implying the presence of partially supporting
solutions. Interestingly, $\frac{D}{2}<k\le D$ are excluded, indicating
that the maximal face dimension of the convex hull is $\frac{D}{2}$.
Each face is the convex hull of a set of $k\le\frac{D}{2}$ points
where each point resides in the $2D$ subspace spanned by a pair of
(real and imaginary) Fourier harmonics. The ring manifolds are closely
related to the \emph{trigonometric moment curve}, whose convex hull
geometrical properties have been extensively studied \citep{kye2013faces,smilansky1985convex}.

In conclusion, the smoothness of the convex hulls becomes apparent
in the distinct patterns of the support dimensions (compare Figs.
\ref{fig:Ellipsoids-with-Realistic}(d) and \ref{fig:Ori_Uniform}(e)).
However, we see that all manifolds with $D_{\text{g}}\thicksim5$
or larger share common trends. As the size of the manifold increases,
the capacity, and geometry vary smoothly, exhibiting a smooth cross-over
from high capacity with low radius and dimension, to low capacity
with large radius and dimension. This crossover occurs as $R_{\text{g}}\propto1/\sqrt{D}_{g}$.
Also, our examples demonstrate, in many cases when the size is smaller
than the crossover value, the manifold dimensionality is substantially
smaller than that expected from naive second order statistics, highlighting
the saliency and significance of our anchor geometry.

\section{Manifolds with Sparse Labels\label{sec:Sparse-Labels}}

So far, we have assumed that the number of manifolds with positive
labels is approximately equal to the number of manifolds with negative
labels. In this section, we consider the case where the two classes
are unbalanced such that the number of positively-labeled manifolds
is far less than the negatively-labeled manifolds (the opposite scenario
is equivalent). This is a special case of the problem of classification
of manifolds with heterogenous statistics, where manifolds have different
geometries or label statistics. We begin by addressing the capacity
of mixtures of manifolds and then focus on sparsely labeled manifolds.

\subsection{Mixtures of manifold geometries \label{subsec:Ensembles}}

Our theory of manifold classification is readily extended to heterogeneous
ensemble of manifolds, consisting of $L$ distinct classes. In the
replica theory, the shape of the manifolds appear only in the free
energy term, $G_{1}$ (see Appendix, Eq. \eqref{eq:G1-1}). For a
mixture statistics, the combined free energy is given by simply averaging
the individual free energy terms over each class $l$. Recall that
this free energy term determines the capacity for each shape, giving
an individual inverse critical load $\alpha_{l}^{-1}$. The inverse
capacity of the heterogeneous mixture is then,

\begin{equation}
\alpha^{-1}=\left\langle \alpha_{l}^{-1}\right\rangle _{l}\label{eq:CapacityMixture}
\end{equation}

where the average is over the fractional proportions of the different
manifold classes. This remarkably simple but generic theoretical result
enables analyzing diverse manifold classification problems, consisting
of mixtures of manifolds with varying dimensions, shapes and sizes.

Eq. \eqref{eq:CapacityMixture} is adequate for classes that differ
in their geometry independent of the assigned labels. In general,
classes may differ in the label statistics (as in the sparse case
studied below) or in geometry that is correlated with labels. For
instance, the positively labelled manifolds may consist of one geometry
and the negatively labelled manifolds may have a different geometry.
How do structural differences between the two classes affect the capacity
of the linear classification? A linear classifier can take advantage
of these correlations by adding \emph{a non-zero bias}. Previously,
it was assumed that the optimal separating hyperplane passes through
the origin; this is reasonable when the two classes are statistically
the same. However, when there are statistical differences between
the label assignments of the two classes, Eq. \eqref{eq:linearSep}
should be replaced by $y^{\mu}(\mathbf{w}\cdot\mathbf{x}^{\mu}-b)\geq\kappa$
where the bias $b$ is chosen to maximize the mixture capacity, \ref{eq:CapacityMixture}.
The effect of optimizing the bias is discussed in detail in the next
section for the sparse labels and in other scenarios in SM (Sec. S3).

\subsection{Manifolds with sparse labels }

We define the sparsity parameter $f$ as the fraction of positively-labeled
manifolds so that $f=0.5$ corresponds to having balanced labels.
From the theory of the classification of a finite set of random points,
it is known that having sparse labels with $f\ll0.5$ can drastically
increase the capacity \citep{gardner1988space}. In this section,
we investigate how the sparsity of manifold labels improves the manifold
classification capacity.

If the separating hyperplane is constrained to go through origin and
the distribution of inputs is symmetric around the origin, the labeling
$y^{\mu}$ is immaterial to the capacity. Thus, the effect of sparse
labels is closely tied to having a non-zero bias. We thus consider
inequality constraints of the form $y^{\mu}(\mathbf{w}\cdot\mathbf{x}^{\mu}-b)\geq\kappa$,
and define the \emph{bias-dependent} capacity of general manifolds
with label sparsity $f$, margin $\kappa$ and bias $b$, as $\alpha_{\text{M}}(\kappa,f,b)$.
Next, we observe that the bias acts as a positive contribution to
the margin for the positively-labeled population and as a negative
contribution to the negatively-labeled population. Thus, the dependence
of $\alpha_{\text{M}}(\kappa,f,b)$ on both $f$ and $b$ can be expressed
as, 
\begin{equation}
\alpha_{\text{M}}^{-1}(\kappa,f,b)\equiv f\alpha_{\text{M}}^{-1}\left(\kappa+b\right)+(1-f)\alpha_{\text{M}}^{-1}\left(\kappa-b\right)\label{eq:alpha_sparse_M}
\end{equation}
where $\alpha_{\text{M}}(x)$ is the classification capacity \emph{with
zero bias }(and hence equivalent to the capacity with $f=0.5$) for
the same manifolds. Note that Eq. \eqref{eq:alpha_sparse_M} is similar
to Eq. \eqref{eq:CapacityMixture} for mixtures of manifolds. The
actual capacity with sparse labels is given by optimizing the above
expression with respect to $b$, i.e., 
\begin{equation}
\alpha_{\text{M}}(\kappa,f)=\mbox{max}_{b}\alpha_{\text{M}}(\kappa,f,b)\label{eq:alphaSparseMax}
\end{equation}

In the following, we consider for simplicity the effect of sparsity
for zero margin, $\kappa=0$.

Importantly, if $D$ is not large, the effect of the manifold geometry
in sparsely labeled manifolds can be much larger than that for non-sparse
labels. For non-sparse labels, the capacity ranges between $2$ and
$(D+0.5)^{-1}$, while for sparse manifolds the upper bound can be
much larger. Indeed, small-sized manifolds are expected to have capacity
that increases upon decreasing $f$ as $\alpha_{\text{M}}(0,f)\propto\frac{1}{f\left|\log f\right|}$,
similar to $P$ uncorrelated points \citep{gardner1988space}. This
potential increase in capacity of sparse labels is however strongly
constrained by the manifold size, since when the manifolds are large,
the solution has to be orthogonal to the manifold directions so that
$\alpha_{\text{M}}(0,f)\approx\frac{1}{D}$. Thus, the geometry of
the manifolds plays an important role in controlling the effect of
sparse labels on capacity. These aspects are already seen in the case
of sparsely labeled $\ell_{2}$ balls (Appendix \ref{sec:AppendixBsparse}).
Here we summarize the main results for general manifolds.

\textbf{Sparsity and size:} There is a complex interplay between label
sparsity and manifold size. Our analysis yields three qualitatively
different regimes:

\emph{Low $R_{\text{g}}$}: When the Gaussian radius of the manifolds
are small, i.e., \textbf{$R_{\text{g}}<1$}, the extent of the manifolds
is noticeable only when the dimension is high. Similar to our previous
analysis of high dimensional manifolds, we find here that the sparse
capacity is equivalent to the capacity of sparsely labeled random
\emph{points} with an effective margin given by $R_{\text{g}}\sqrt{D_{\text{g}}}$,
\begin{equation}
\alpha_{\text{M}}(f)\approx\alpha_{0}(f,\kappa=R_{\text{g}}\sqrt{D_{\text{g}}})\label{eq:aca0-2}
\end{equation}
where $\alpha_{0}(\kappa,f)=\mbox{max}_{b}\alpha_{0}(\kappa,f,b)$
from the Gardner theory. It should be noted that $\kappa_{\text{g}}$
in the above equation has a noticeable effect only for moderate sparsity.
It has a negligible effect when $f\rightarrow0,$ since the bias is
large and dominates over the margin.

\emph{Moderate sizes, $R_{\text{g}}>2$}: In this case, the equivalence
to the capacity of points breaks down. Remarkably we find that the
capacity of general manifolds with substantial size is well approximated
by that of equivalent $\ell_{2}$ balls with the same sparsity $f$
and with dimension and radius equal to the Gaussian dimension and
radius of the manifolds, namely

\begin{equation}
\alpha_{\text{M}}(f)\approx\alpha_{\text{Ball}}(f,R_{\text{g}},D_{\text{g}})\label{eq:alMsparse}
\end{equation}
Surprisingly, unlike the nonsparse approximation, where the equivalence
of general manifold to balls, Eq \ref{eq:capacityLargeD-1}, is valid
only for high dimensional manifolds, in the sparse limit, the spherical
approximation is not restricted to large $D$. Another interesting
result is that the relevant statistics are given by the Gaussian geometry,
$R_{\text{g}}$and $D_{\text{g}}$, even when $R_{\text{g}}$ is not
small. The reason is that for small $f$, the bias is large. In that
case, the positively labeled manifolds have large positive margin
$b$ and are fully supporting giving a contribution to the inverse
capacity of $b^{2}$ regardless of their detailed geometry. On the
other hand, the negatively labeled manifolds have large negative margin,
implying that most of them are far from the separating plane (interior)
and a small fraction have touching support. The fully supporting configurations
have negligible probability, hence the overall geometry is well approximated
by the Gaussian quantities $D_{\text{g}}$ and $R_{\text{g}}$.

\textbf{Scaling relationship between sparsity and size: }A further
analysis of the capacity of balls with sparse labels shows it retains
a simple form $\alpha(\bar{f},D)$ (see Appendix \ref{sec:AppendixBsparse}
Eq. \ref{eq:alfbarx}) which depends on $R$ and $f$ only through
the scaled sparsity $\bar{f}=fR^{2}$. The reason for the scaling
of $f$ with $R^{2}$ is as follows. When the labels are sparse, the
dominant contribution to the inverse capacity comes from the minority
class, and so the capacity is $\alpha_{\text{Ball}}^{-1}\approx fb^{2}$
for large $b$. On the other hand, the optimal value of $b$ depends
on the balance between the contributions from both classes and scales
linearly with $R$, as it needs to overcome the local fields from
the spheres. Thus, $\alpha^{-1}\propto fR^{2}$ .

Combining Eq. \ref{eq:alfbarx} with Eq. \ref{eq:alMsparse} yields
for general sparsely labeled manifolds,

\begin{equation}
\alpha_{\text{M}}^{-1}=\bar{f}\bar{b}^{2}+\int_{\bar{b}}^{\infty}dt\chi_{D_{\text{g}}}(t)(t-\bar{b})^{2}\label{eq:alMsparse-1}
\end{equation}

where scaled sparsity is $\bar{f}=f(1+R_{\text{g}}^{2})\lesssim1$
.

Note we have defined the \emph{scaled sparsity} $\bar{f}=f(1+R_{\text{g}}^{2})$
rather than $fR^{2}$to yield a smoother cross-over to the small $R_{\text{g}}$
regime. Similarly, we define the optimal scaled bias as $\bar{b}=b\sqrt{1+R_{\text{g}}^{2}}\gg1.$
Qualitatively, the function $\alpha$ is roughly proportional to $\bar{f}^{-1}$
with a proportionality constant that depends on $D_{\text{g}}$. In
the extreme limit of $\left|\log\bar{f}\right|\gg D_{\text{g}}$,
we obtain the sparse limit $\alpha\propto\left|\bar{f}\log\bar{f}\right|^{-1}$.
Note that if $f$ is sufficiently small, the gain in capacity due
to sparsity occurs even for large manifolds as long as $\bar{f}<1$.

\emph{Large $R_{\text{g}}$ regime: $\bar{f}>1$:} Finally, when $R_{\text{g}}$
is sufficiently large such that $\bar{f}$ increases above $1$, $\bar{b}$
is of order $1$ or smaller, and the capacity is small with a value
that depends on the detailed geometry of the manifold. In particular,
when $\bar{f}\gg1$, the capacity of the manifold approaches $\alpha_{\text{M}}(f)\rightarrow\frac{1}{D}$,
not $\frac{1}{D_{\text{g}}}$.

\begin{figure}
\begin{centering}
\includegraphics[width=1\columnwidth]{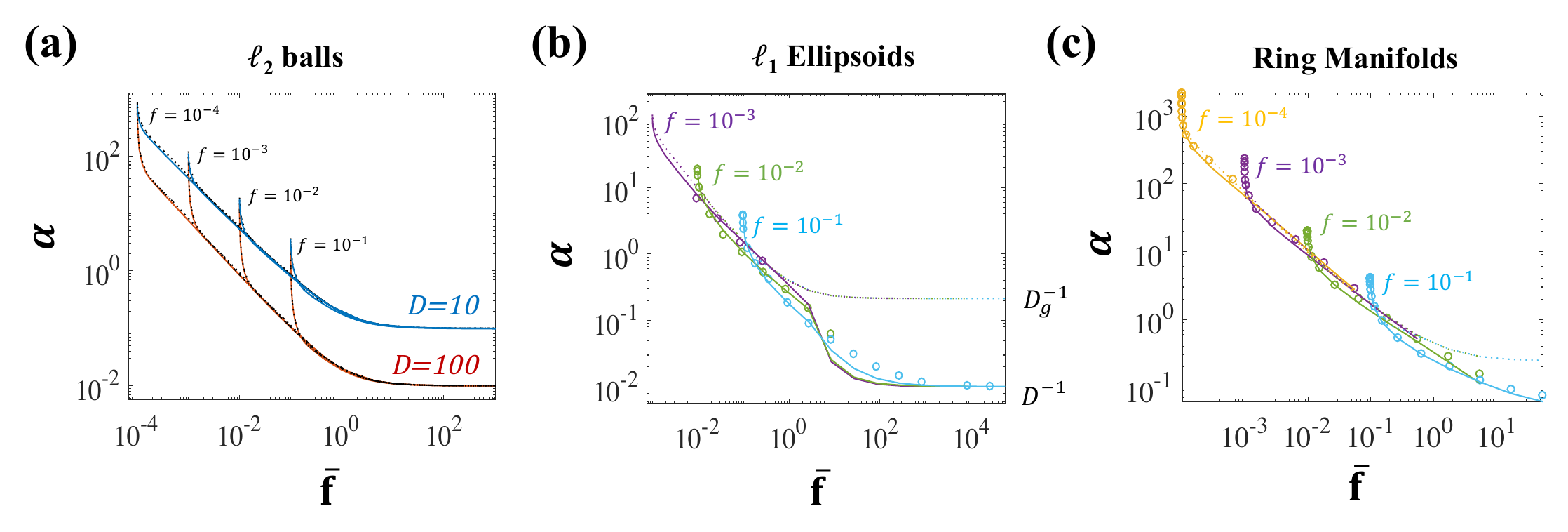} 
\par\end{centering}
\caption{\textcolor{black}{(a-b) Classification of $\ell_{2}$ balls with sparse
labels: (a) Capacity of $\ell_{2}$ balls as a function of $\bar{f}=f(1+r^{2})$,
for $D=100$ (red) and $D=10$ (blue): (solid lines) mean field theory,
(dotted lines) is an approximation interpolating between }Eqs. \ref{eq:aca0-2}
and \ref{eq:alMsparse-1}\textcolor{black}{{} (details in SM, Sec.
S9). (b--c) Classification of general manifolds with sparse labels:}\textbf{\textcolor{black}{{}
}}\textcolor{black}{(b) Capacity of $\ell_{1}$ ellipsoids with $D=100$,
where the first 10 components are equal to $r$, and the remaining
90 components are $\frac{1}{2}r$, as a function of $\bar{f}=(1+R_{\text{g}}^{2})$.
$r$ is varied from $10^{-3}$ to $10^{3}$: (circles) numerical simulations,
(lines) mean field theory, (dotted lines) spherical approximation.
(c) Capacity of ring manifolds with a Gaussian fall-off spectrum with
$\sigma=0.1$ and $D=100$ (; details in SM).} Fig. \ref{fig:SparseManifolds}(c)
shows the capacity of ring manifolds whose Fourier components have
a Gaussian fall-off, i.e. $R_{n}=A\exp\left[-\frac{1}{2}\left(2\pi(n-1)\sigma\right)^{2}\right]$.
\textcolor{black}{\label{fig:SparseManifolds}}}
\end{figure}

To demonstrate these remarkable predictions, Eq. \eqref{eq:alMsparse},
we present in Fig. \ref{fig:SparseManifolds} the capacity of three
classes of sparsely labeled manifolds:$\ell_{2}$ balls, $\ell_{1}$
ellipsoids and ring manifolds. In all cases, we show the results of
numerical simulations of the capacity, the full mean field solution,
and the spherical approximation, \ref{eq:alMsparse}, across several
orders of magnitude of sparsity and size as a function of the scaled
sparsity $\bar{f}=f(1+r^{2})$. In each example, there is a good agreement
between the three calculations for the range of $\bar{f}<1$. Furthermore,
the drop in $\alpha$ with increasing $\bar{f}$ is similar in all
cases, except for an overall vertical shift which is due to the different
$D_{\text{g}}$, similar to the effect of dimension in $\ell_{2}$
balls $(a)$. In the regime of moderate radii, results for different
$f$ and $r$ all fall on a universal curve which depends only on
$\bar{f}$, as predicted by the theory. For small $r$, the capacity
deviates from this scaling as it is dominated by $f$ alone, similar
to sparsely labeled points. When $\bar{f}>1$, the curves deviate
from the spherical approximation. The true capacity (as revealed by
simulations and full mean field) rapidly decreases with $\bar{f}$
and saturates at $\frac{1}{D}$, rather than to the $\frac{1}{D_{\text{g}}}$
limit of the spherical approximation. Finally, we note that our choice
of parameters in (c) (SM Section 7) were such that $R_{\text{g}}$
(entering in the scaled sparsity) was significantly different from
simply an average radius. Thus, the agreement with the theory illustrates
the important role of the Gaussian geometry in the sparse case.

\begin{figure}[h]
\includegraphics[width=1\columnwidth]{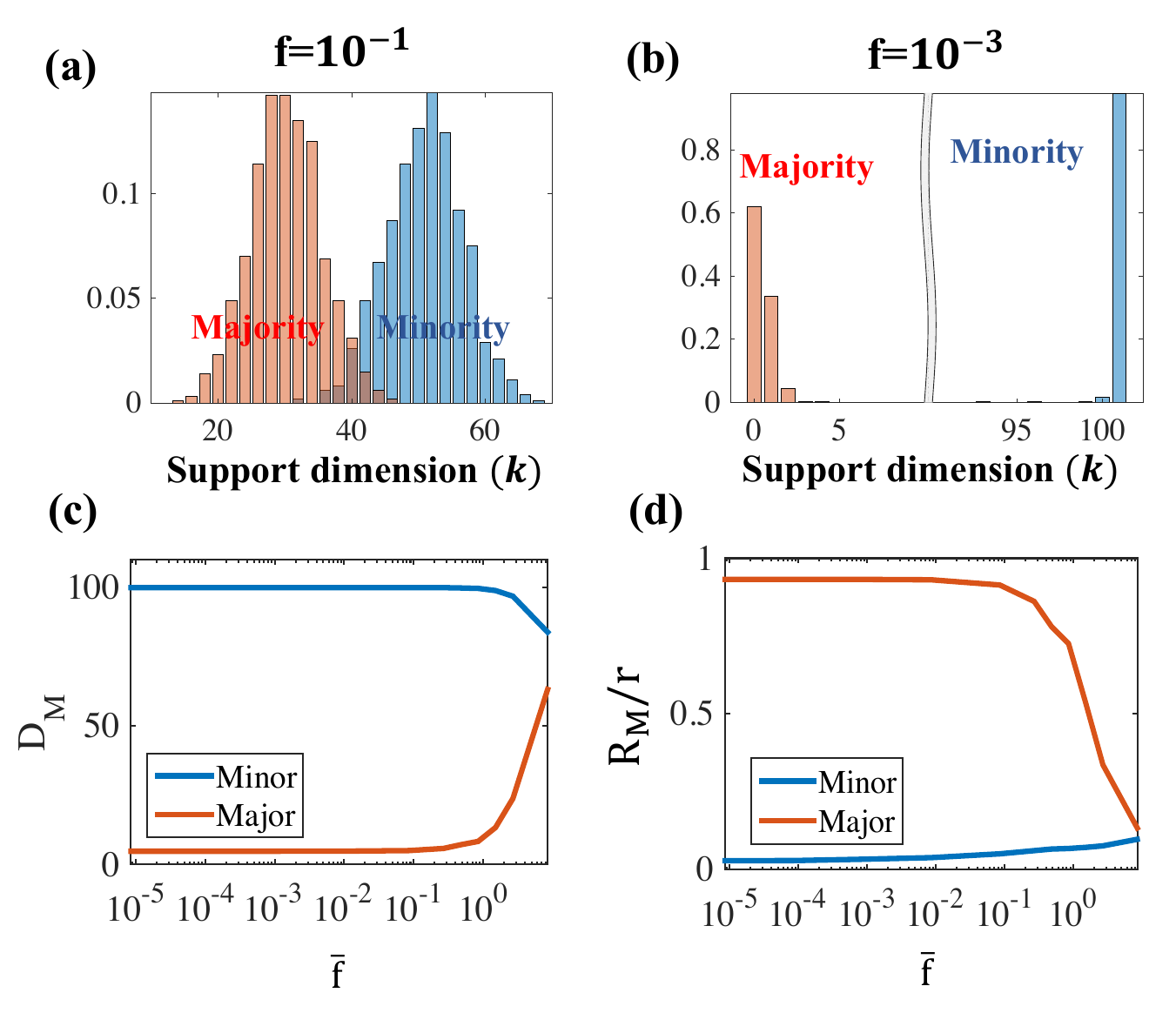}

\caption{\textcolor{black}{Manifold configurations and geometries for classification
of $\ell_{1}$ ellipsoids with sparse labels, analyzed separately
in terms of the majority and minority classes. The radii for $\ell_{1}$
ellipsoids are $R_{i}=r$ for $1\le i\le10$ and $R_{i}=\frac{1}{2}r$
for $11\le i\le100$ with $r=10$. (a) Histogram of support dimensions
for moderate sparsity $f=0.1$: (blue) minority and (red) majority
manifolds. (b) Histogram of support dimensions for high sparsity $f=10^{-3}$:
(blue) minority and (red) majority manifolds. (c) Manifold dimension
as a function of $\bar{f}=f(1+R_{\text{g}}^{2})$ as $f$ is varied:
(blue) minority and (red) majority manifolds. (d) Manifold radius
relative to the scaling factor $r$ as a function of $\bar{f}$: (blue)
minority and (red) majority manifolds.\label{fig:Sparse_k}}}
\end{figure}

As discussed above, in the (high and moderate) sparse regimes, a large
bias alters the geometry of the two classes in different ways. To
illustrate this important aspect, we show in Fig. \ref{fig:Sparse_k},
the effect of sparsity and bias on the geometry of the $\ell_{1}$ellipsoids
studied in Fig. \ref{fig:SparseManifolds}(c). Here we show the evolution
of $R_{\text{M}}$ and $D_{\text{M}}$ for the majority and minority
classes as $\bar{f}$ increases. Note that despite the fact that the
shape of the manifolds is the same for all $f,$ their\emph{ manifold
anchor geometry} depends on both class membership and sparsity levels
because these measures depend on the margin. When $\bar{f}$ is small,
the minority class has $D_{\text{M}}=D$ as seen in Fig. \ref{fig:Sparse_k}(c),
i.e., the minority class manifolds are close to be fully supporting
due to the large positive margin. This can also be seen in the distributions
of support dimension shown in Fig.\ref{fig:Sparse_k}(a)-(b). On the
other hand, the majority class has $D_{\text{M}}\approx D_{\text{g}}=2\text{\ensuremath{\log}}D_{1}\ll D$,
and these manifolds are mostly in the interior regime. As $f$ increases,
the geometrical statistics for the two classes become more similar.
This is seen in Fig.\ref{fig:Sparse_k}(c)-(d) where $D_{\text{M}}$
and $R_{\text{M}}$ for both majority and minority classes converge
to the zero margin value for large $\bar{f}=f\left(1+R_{\text{g}}^{2}\right)$.

\section{Summary and Discussion\label{sec:Summary}}

\textbf{Summary:} We have developed a statistical mechanical theory
for linear classification of inputs organized in perceptual manifolds,
where all points in a manifold share the same label. The notion of
perceptual manifolds is critical in a variety of contexts in computational
neuroscience modeling and in signal processing. Our theory is not
restricted to manifolds with smooth or regular geometries; it applies
to any compact subset of a $D$-dimensional affine subspace. Thus,
the theory is applicable to manifolds arising from any variation in
neuronal responses with a continuously varying physical variable,
or from any sampled set arising from experimental measurements on
a limited number of stimuli.

The theory describes the capacity of a linear classifier to separate
a dichotomy of general manifolds (with a given margin) by a universal
set of mean field equations. These equations may be solved analytically
for simple geometries, but for more complex geometries, we have developed
iterative algorithms to solve the self-consistent equations. The algorithms
are efficient and converge fast as they involve only solving for $O(D)$
variables of a single manifold, rather than invoking simulations of
a full system of $P$ manifolds embedded in $\mathbb{R}^{N}$.

\textbf{Applications:}

The statistical mechanical theory of perceptron learning has long
provided a basis for understanding the performance and fundamental
limitations of single layer neural architectures and their kernel
extensions. However, the previous theory only considered a finite
number of random points with no underlying geometric structure, and
could not explain the performance of linear classifiers on large,
possibly infinite number of inputs organized as distinct manifolds
by the variability due to changes in physical parameters of objects.
The theory presented in this work can explain the capacity and limitations
of linear classification of general manifolds.

This new theory is important for the understanding of how sensory
neural systems perform invariant perceptual discrimination and recognition
tasks of realistic stimuli. Furthermore, beyond estimating the classification
capacity, our theory provides theoretically based geometric measures
for assessing the quality of the neural representations of the perceptual
manifolds. There are a variety of ways for defining the the geometry
of manifolds. Our geometric measures are unique in that they determine
the ability to linearly separate the manifold, as our theory shows.

Our theory focuses on linear classification. However, it has broad
implications for non-linear systems, and in particular for deep networks.
First, most models of sensory discrimination and recognition in biological
and artificial deep architectures model the readouts of the networks
as linear classifiers operating on the top sensory layers. Thus, our
manifold classification capacity and geometry can be applied to understand
the performance of the deep network. Furthermore, the computational
advantage of having multiple intermediate layers can be assessed by
comparing the performance of a hypothetical linear classifier operating
on these layers. In addition, the changes in the quality of representations
across the deep layers can be assessed by comparing the changes in
the manifold's geometries across layers. Indeed, previous discussions
of sensory processing in deep networks hypothesized that neural object
representations become increasingly untangled as the signal propagates
along the sensory hierarchy. However, no concrete measure of untangling
has been provided. The geometric measures derived from our theory
can be used to quantify the degree of entanglement, tracking how the
perceptual manifolds are nonlinearly reformatted as they propagate
through the multiple layers of a neural network to eventually allow
for linear classification in the top layer. Notably, the statistical,
population-based, nature of our geometric measures renders them ideally
suited for comparison between layers of different sizes and nonlinearities,
as well as between different deep networks or between artificial networks
and biological ones. Lastly, our theory can suggest new algorithms
for building deep networks, for instance by imposing successive reduction
of manifold dimensions and radii as part of an unsupervised learning
strategy.

We have discussed above the domain of visual object classification
and recognition, which has received immense attention in recent years.
However, we would like to emphasize that our theory can be applied
for modeling neural sensory processing tasks in other modalities.
For instance, it can be used to provide insight on how the olfactory
system performs discrimination and recognition of odor identity in
the presence of orders of magnitude variations in odor concentrations.
Neuronal responses to sensory signals are not static but vary in time.
Our theory can be applied to explain how the brain correctly decodes
the stimulus identity despite its temporal non-stationarity. 

Some of these applications may require further extensions of the present
theory. The most important ones (currently the subject of ongoing
work) include:

\textbf{Correlations: }In the present work we have assumed that the
directions of the affine subspaces of the different manifolds are
uncorrelated. In realistic situations we expect to see correlations
in the manifold geometries, mainly of two types: One is center-center
correlations. Such correlations can be harmful for linear separability
\citep{monasson1992properties,lopez1995storage}. Another is correlated
variability in which the directions of the affine subspaces are correlated
but not the centers. Positive correlations of the latter form are
beneficial for separability. In the extreme case when the manifolds
share a common affine subspace, the rank of the union of the subspaces
is $D_{tot}=D$ rather than $D_{tot}=PD$, and the solution weight
vector need only lie in the null space of this smaller subspace. Further
work is needed to extend the present theory to incorporate more general
correlations.

\textbf{Generalization performance:} We have studied the separability
of manifolds with known geometries. In many realistic problems, this
information is not readily available and only samples reflecting the
natural variability of input patterns are provided. These samples
can be used to estimate the underlying manifold model (using manifold
learning techniques \citep{roweis2000nonlinear,tenenbaum2000global})
and/or to train a classifier based upon a finite training set. Generalization
error describes how well a classifier trained on a finite number of
samples would perform on other test points drawn from the manifolds
\citep{amari1992four}. It would be important to extend our theory
to calculate the expected generalization error achieved by the maximum
margin solution trained on \emph{point cloud manifolds}, as a function
of the size of the training set and the geometry of the underlying
full manifolds.

\textbf{Unrealizable classification: }Throughout the present work,
we have assumed that the manifolds are separable by a linear classifier.
In realistic problems, the load may be above the capacity for linear
separation, i.e. $\alpha>\alpha_{\text{M}}(\kappa=0)$. Alternatively,
neural noise may cause the manifolds to be unbounded in extent, with
the tails of their distribution overlapping so that they are not separable
with zero error. There are several ways to handle this issue in supervised
learning problems. One possibility is to nonlinearly map the unrealizable
inputs to a higher dimensional feature space, via a multi-layer network
or nonlinear kernel function, where the classification can be performed
with zero error. The design of multilayer networks could be facilitated
using manifold processing principles uncovered by our theory.

Another possibility is to introduce an optimization problem allowing
a small training error, for example, using an SVM with complementary
slack variables \citep{vapnik1998statistical}. These procedures raise
interesting theoretical challenges, including understanding how the
geometry of manifolds change as they undergo nonlinear transformations,
as well as investigating by statistical mechanics, the performance
of a linear classifier of manifolds with slack variables \citep{risau2001statistical}.

In conclusion, we believe the application of this theory and its corollary
extensions will precipitate novel insights into how perceptual systems,
biological or artificial, efficiently code and process complex sensory
information.
\begin{acknowledgments}
We would like to thank Uri Cohen, Ryan Adams, Leslie Valiant, David
Cox, Jim DiCarlo, Doris Tsao, and Yoram Burak for helpful discussions.
The work is partially supported by the Gatsby Charitable Foundation,
the Swartz Foundation, the Simons Foundation (SCGB Grant No. 325207),
the NIH, and the Human Frontier Science Program (Project RGP0015/2013).
D. D. Lee also acknowledges the support of the US National Science
Foundation, Army Research Laboratory, Office of Naval Research, Air
Force Office of Scientific Research, and Department of Transportation. 
\end{acknowledgments}

\appendix

\section{Replica Theory of Manifold Capacity\label{sec:AppendixReplica-1}}

In this section, we outline the derivation of the mean field replica
theory summarized in Eqs. \eqref{eq:inv_capacity}-\ref{eq:kappaG}.
We define the capacity of linear classification of manifolds, $\alpha_{\text{M}}(\kappa)$,
as the maximal load, $\alpha=\frac{P}{N}$, for which with high probability
a solution to $y^{\mu}\mathbf{w}\cdot\mathbf{x}^{\mu}\geq\kappa$
exists for a given $\kappa$. Here $\mathbf{x}^{\mu}$ are points
on the $P$ manifolds $M^{\mu},$ Eq. \eqref{eq:d+1manifolds}, and
we assume that all $NP(D+1)$ components of $\left\{ \mathbf{u}_{i}^{\mu}\right\} $
are drawn independently from a Gaussian distribution with zero mean
and variance $\frac{1}{N}$, and that the binary labels $y^{\mu}=\pm1$
are randomly assigned to each manifold with equal probabilities. We
consider the thermodynamic limit where $N,\,P\rightarrow\infty$ but
$\alpha=\frac{P}{N}$, and $D$ are finite.

Note that the geometric margin, $\kappa^{\prime}$, defined as the
distance from the solution hyperplane is given by $y^{\mu}\mathbf{w}\cdot\mathbf{x}^{\mu}\geq\kappa^{\prime}\left\Vert \mathbf{w}\right\Vert =\kappa^{\prime}\sqrt{N}$.
However, this distance depends on the scale of the input vectors $\mathbf{x}^{\mu}$.
The correct scaling of the margin in the thermodynamic limit is $\kappa^{\prime}=\frac{\left\Vert \mathbf{x}\right\Vert }{\sqrt{N}}\kappa$.
Since we adopted the normalization of $\left\Vert \mathbf{x}^{\mu}\right\Vert =O(1)$,
the correct scaling of the margin is $y^{\mu}\mathbf{w}\cdot\mathbf{x}^{\mu}\geq\kappa$
.

\textbf{Evaluation of solution volume: }Following Gardner's replica
framework, we first consider the volume $Z$ of the solution space
for $\alpha<\alpha_{\text{M}}(\kappa)$ . We define the signed projections
of the the $i$th direction vector $\mathbf{u}_{i}^{\mu}$ on the
solution weight as $H_{i}^{\mu}=\sqrt{N}y^{\mu}\mathbf{w}\cdot\mathbf{u}_{i}^{\mu}$,
where $i=1,...,D+1$ and $\mu=1,...,P$. Then, the separability constraints
can be written as $\sum_{i=1}^{D+1}S_{i}H_{i}^{\mu}\geq\kappa$ .
Hence the volume can be written as 
\begin{equation}
Z=\int d^{N}\mathbf{w}\delta(\mathbf{w}^{2}-N)\,\Pi_{\mu=1}^{P}\Theta_{\mu}\left(g_{\mathcal{S}}(\vec{H}^{\mu})-\kappa\right)\label{eq:V-1-1}
\end{equation}
where $\Theta(x)$ is a Heaviside step function. $g_{\mathcal{S}}$
is the \emph{support function} of ${\cal \mathcal{S}}$ defined for
Eq. \ref{eq:kappaG} as $g_{\mathcal{S}}(\vec{V})=\min_{\vec{S}}\left\{ \vec{V}\cdot\vec{S}\mid\vec{S}\in{\cal \mathcal{S}}\right\} $.

The volume defined above depends on the the quenched random variables
$\mathbf{u}_{i}^{\mu}$ and $y^{\mu}$ through $H_{i}^{\mu}$. It
is well known that in order to obtain the typical behavior in the
thermodynamic limit, we need to average $\log Z$, which we carry
out using the replica trick, $\langle\log Z\rangle=\lim_{n\rightarrow0}\frac{\langle Z^{n}\rangle-1}{n}$,
where $\langle\rangle$ refers to the average over $\mathbf{u}_{i}^{\mu}$
and $y^{\mu}$. For natural $n,$ we need to evaluate, 
\begin{align}
\langle Z^{n}\rangle & =\int\prod_{\alpha}^{n}d\mathbf{w}_{\alpha}\delta(\mathbf{w}_{\alpha}^{2}-N)\prod_{\mu}^{P}\int\mathbb{D}\vec{H}{}^{\mu\alpha}\label{eq:Vn0-1-1}\\
 & \langle\prod_{i}^{D+1}\sqrt{2\pi}\delta(H_{i}^{\mu\alpha}-y^{\mu}w_{\alpha}^{T}\mathbf{u}_{i}^{\mu})\rangle_{\mathbf{u}_{i}^{\mu},y^{\mu}}\nonumber 
\end{align}
where we have used the notation, 
\begin{equation}
\mathbb{D}\vec{H}=\Pi_{i=1}^{D+1}\frac{dH_{i}}{\sqrt{2\pi}}\Theta\left(g_{\mathcal{S}}(\vec{H})-\kappa\right)
\end{equation}

Using Fourier representation of the delta functions, we obtain 
\begin{align}
\langle Z^{n}\rangle & =\int\prod_{\alpha}^{n}d\mathbf{w}_{\alpha}\delta(\mathbf{w}_{\alpha}^{2}-N)\prod_{\mu}^{P}\int\mathbb{D}\vec{H}{}^{\mu\alpha}\label{eq:Vn0-2-1}\\
 & \prod_{i=1}^{D+1}\int\frac{d\hat{H}_{i}^{\mu\alpha}}{\sqrt{2\pi}}\left\langle \text{exp}\left\{ i\hat{H}_{i}^{\mu\alpha}(H_{i}^{\mu\alpha}-y^{\mu}\mathbf{w}_{\alpha}^{T}\mathbf{u}_{i}^{\mu})\right\} \right\rangle _{\mathbf{u}_{i}^{\mu},y^{\mu}}\nonumber 
\end{align}

Performing the average over the Gaussian distribution of $\mathbf{u}_{i}^{\mu}$
(each of the $N$ components has zero mean and variance $\frac{1}{N}$)
yields, 
\begin{align}
 & \left\langle \text{exp}\sum_{i=1}^{D+1}\sum_{\mu\alpha}\left[i\hat{H}_{i}^{\mu\alpha}(-y^{\mu}\sum_{j=1}^{N}w_{\alpha}^{j}\mathbf{u}_{i,j}^{\mu})\right]\right\rangle _{\mathbf{u}_{i}^{\mu},y^{\mu}}\\
 & =\text{exp}\left\{ -\frac{1}{2}\sum_{\alpha\beta}\mathfrak{q}_{\alpha\beta}\sum_{i\mu}\hat{H}_{i}^{\mu\alpha}\hat{H}_{i}^{\mu\beta}\right\} \nonumber 
\end{align}
where, $\mathfrak{q}_{\alpha\beta}=\frac{1}{N}\sum_{j=1}^{N}w_{\alpha}^{j}w_{\beta}^{j}$.
Thus, integrating the variables $\hat{H}_{i}^{\mu\alpha}$ yields
\begin{align}
\langle Z^{n}\rangle & =\int\prod_{\alpha=1}^{n}d\mathbf{w}_{\alpha}\delta(\mathbf{w}_{\alpha}^{2}-N)\int d\mathfrak{q}_{\alpha\beta}\Pi_{\alpha\beta}\label{eq:Vn-1}\\
 & \cdot\delta\left(N\mathfrak{q}_{\alpha\beta}-\mathbf{w}_{\alpha}^{T}\mathbf{w}_{\beta}\right)\left[\exp\left(-\frac{(D+1)}{2}\text{log}\text{det}\mathfrak{q}\right)X\right]^{P}\nonumber 
\end{align}
where 
\begin{align}
X & =\int\prod_{\alpha}\mathbb{D}\vec{H}^{\alpha}\exp\left[-\frac{1}{2}\sum_{i,\alpha,\beta}H_{i}^{\alpha}(\mathfrak{q}^{-1})_{\alpha\beta}H_{i}^{\beta}\right]\label{eq:SquareBracket-1}
\end{align}
and we have used the fact that all manifolds contribute the same factor.

We proceed by making the replica symmetric ansatz on the order parameter
$\mathfrak{q}_{\alpha\beta}$ at its saddle point, $\mathfrak{q}_{\alpha\beta}=(1-q)\delta_{\alpha\beta}+q$,
from which one obtains in the $n\rightarrow0$ limit: 
\begin{equation}
\mathfrak{q}_{\alpha\beta}^{-1}=\frac{1}{1-q}\delta_{\alpha\beta}-\frac{q}{(1-q)^{2}}
\end{equation}
and 
\begin{equation}
\text{log}\text{det}\mathfrak{q}=n\log(1-q)+\frac{nq}{1-q}\label{eq:logdetq-1}
\end{equation}

Thus the exponential term in $X$ can be written as 
\begin{equation}
\exp\left[-\frac{1}{2}\sum_{\alpha i}\frac{\left(H_{i}^{\alpha}\right)^{2}}{1-q}+\frac{1}{2}\sum_{i}\left(\frac{\sqrt{q}}{1-q}\sum_{\alpha}H_{i}^{\alpha}\right)^{2}\right]
\end{equation}

Using the Hubbard--Stratonovich transformation, we obtain 
\begin{equation}
X=\int D\vec{T}\left[\int\mathbb{D}\vec{H}\exp\left\{ -\frac{1}{2}\frac{\vec{H}^{2}}{1-q}+\frac{\sqrt{q}}{1-q}\vec{H}\cdot\vec{T}\right\} \right]^{n}
\end{equation}
where $D\vec{T}=\Pi_{i}\frac{dT_{i}}{\sqrt{2\pi}}\exp\left(-\frac{T_{i}^{2}}{2}\right)$.
Completing the square in the exponential and using $\int D\vec{T}A^{n}=\exp n\int D\vec{T}\log A$
in the $n\rightarrow0$ limit, we obtain, $X=\exp\left(\frac{nq(D+1)}{2(1-q)}+n\int D\vec{T}\log z(\vec{T})\right)$
with 
\begin{equation}
z(\vec{T})=\int\mathbb{D}\vec{H}\exp\left\{ -\frac{1}{2(1-q)}||\vec{H}-\sqrt{q}\vec{T}||^{2}\right\} \label{eq:Zt-1}
\end{equation}

Combining these terms, we write the last factor in Eq. \eqref{eq:Vn-1}
as $\exp nPG_{1}$ where,

\begin{equation}
G_{1}=\int D\vec{T}\log z(\vec{T})-\frac{(D+1)}{2}\log(1-q)\label{eq:G1-1}
\end{equation}

The first factors in $\langle Z^{n}\rangle$, Eq. \eqref{eq:Vn-1},
can be written as $\exp nNG_{0}$, where as in the Gardner theory,
the entropic term in the thermodynamic limit is 
\begin{equation}
G_{0}(q)=\frac{1}{2}\ln(1-q)+\frac{q}{2(1-q)}\label{eq:G0-1}
\end{equation}
and represents the constraints on the volume of $\mathbf{w}_{\alpha}$due
to normalization and the order parameter $\mathfrak{q}$. Combining
the $G_{0}$ and $G_{1}$contributions, we have 
\begin{equation}
\langle Z^{n}\rangle_{t_{0},t}=e^{Nn\left[G_{0}(q)+\alpha G_{1}(q)\right]}
\end{equation}

The classification constraints contribute $\alpha G_{1}$, with Eq.
\eqref{eq:G1-1}, and 
\begin{align}
z(\vec{T}) & =\int\Pi_{i=1}^{D+1}\frac{dY_{i}}{\sqrt{2\pi(1-q)}}\exp\left(-\frac{\vec{Y}^{2}}{2(1-q)}\right)\label{eq:Z_append-1}\\
 & \Theta\left(g_{\mathcal{S}}(\sqrt{q}\vec{T}+\vec{Y})-\kappa\right)\nonumber 
\end{align}
where, we have written the fields $H_{i}$ as 
\begin{equation}
H_{i}=\sqrt{q}T_{i}+Y_{i}\label{eq:h_append-1}
\end{equation}

Note that $\sqrt{q}T_{i}$ represents the quenched random component
due to the randomness in the $\mathbf{u}_{i}^{\mu}$, and $Y_{i}$
is the ``thermal'' component due to the variability within the solution
space. The order parameter $q$ is calculated via $0=\frac{\partial G_{0}}{\partial q}+\alpha\frac{\partial G_{1}}{\partial q}$
.

\textbf{Capacity: }In the limit where $\alpha\rightarrow\alpha_{\text{M}}(\kappa)$
, the overlap between the solutions become unity and the volume shrinks
to zero. It is convenient to define $Q=\frac{q}{1-q}$ and study the
limit of $Q\rightarrow\infty$. In this limit the leading order is
\begin{equation}
\langle\log Z\rangle=\frac{Q}{2}\left[1-\alpha\langle F(\vec{T})\rangle_{\vec{T}}\right]\label{eq:logVLargeQ-1}
\end{equation}
where the first term is the contribution from $G_{0}\rightarrow\frac{Q}{2}$.
The second term comes from $G_{1}\rightarrow-\frac{Q}{2}\alpha\langle F(\vec{T})\rangle_{\vec{T}}$,
where the average is over the Gaussian distribution of the $D+1$
dimensional vector $\vec{T}$, and 
\begin{equation}
F(\vec{T})\rightarrow-\frac{2}{Q}\log z(\vec{T})
\end{equation}
is independent of $Q$ and is given by replacing the integrals in
Eq. \eqref{eq:Z_append-1} by their saddle point, which yields 
\begin{equation}
F(\vec{T})=\min_{\vec{V}}\left\{ \left\Vert \vec{V}-\vec{T}\right\Vert ^{2}\mid g_{\mathcal{S}}\left(\vec{V}\right)-\kappa\geq0\right\} \label{eq:F(t)-2}
\end{equation}

At the capacity, $\log Z$ vanishes, the capacity of a general manifold
with margin $\kappa$, is given by, 
\begin{align}
\alpha_{\text{M}}^{-1}(\kappa) & =\langle F(\vec{T})\rangle_{\vec{T}}\label{eq:alphaLines-1}\\
F(\vec{T}) & =\min_{\vec{V}}\left\{ \left\Vert \vec{V}-\vec{T}\right\Vert ^{2}\mid g_{\mathcal{S}}(\vec{V})-\kappa\geq0\right\} \label{eq:F(t)-1-1}
\end{align}

Finally, we note that the mean squared 'annealed' variability in the
fields due to the entropy of solutions vanishes at the capacity limit,
as $1/Q$ , see Eq. \eqref{eq:Z_append-1} . Thus, the quantity $\left\Vert \vec{V}-\vec{T}\right\Vert ^{2}$in
the above equation represents the annealed variability times $Q$
which remains finite in the limit of $Q\rightarrow\infty$.

\section{Strictly Convex Manifolds\label{sec:AppendixB:StrictlyConvex}}

\subsection*{B.1. General}

Here we evaluate the capacity of strictly convex manifolds, starting
from the expression for general manifolds, Eq. \ref{eq:amMgen}. In
a strictly convex manifold, $\mathcal{S}$ , any point in the line
segment connecting two points $\vec{x}$ and $\vec{y}$, $\vec{x},\vec{y}\in\mathcal{S}$
, other than $\vec{x}$ and $\vec{y}$, belongs to the interior of
$\mathcal{S}$ . Thus, the boundary of the manifold does not contain
edges or flats with spanning dimension $k>1$ except of the entire
manifold, $k=D+1$ . Therefore, there are exactly contributions to
the inverse capacity. When $t_{0}$ obeys $t_{\text{touch}}(\vec{t})>t_{0}-\kappa>t_{\text{fs}}(\vec{t})$,
the integrand of Eq. \ref{eq:amMgen} contributes $\frac{(-\vec{t}\cdot\tilde{s}(\vec{T})-t_{0}+\kappa)^{2}}{1+\left\Vert \tilde{s}(\vec{T})\right\Vert ^{2}}$
, When $t_{0}<\kappa+t_{\text{fs}}(\vec{t})$, the manifold is fully
embedded. In this case, $\vec{v}=0,$ and the integrand reduces to
Eq. \ref{eq:F_full Supp}. In summary, the capacity for convex manifolds
can be written as

\begin{align}
\alpha^{-1}(\kappa)= & \int D\vec{t}\int_{\kappa+t_{\text{fs}}(\vec{t})}^{\kappa+t_{\text{touch}}(\vec{t})}Dt_{0}\frac{(-\vec{t}\cdot\tilde{s}(\vec{t},t_{0})-t_{0}+\kappa)^{2}}{1+\left\Vert \tilde{s}(\vec{t},t_{0})\right\Vert ^{2}}\label{eq:strictConv}\\
 & +\int D\vec{t}\int_{-\infty}^{\kappa+t_{\text{fs}}(\vec{t})}Dt_{0}\left[\left(t_{0}-\kappa\right)^{2}+\left\Vert \vec{t}\right\Vert ^{2}\right]\nonumber 
\end{align}

where $t_{\text{touch}}(\vec{t})$ and $t_{\text{fs}}(\vec{t})$ are
given by Eqs. \ref{eq:tTouch} and \ref{eq:tembedGen}, respectively,
and $\tilde{s}(\vec{t},t_{0})=\arg\min_{\vec{s}}(\vec{v}\cdot\vec{s}),$
with $\vec{v}=\vec{t}+(v_{0}-t_{0})\tilde{s}$.

\subsection*{B.2. $\ell_{2}$ Balls\label{subsec:B.2.AppendixL2Balls}}

In the case of $\ell_{2}$ balls with $D$ and radius $R$, $g(\vec{v})=-R\left\Vert \vec{v}\right\Vert $
. Hence, $t_{\text{touch}}(\vec{t})=R\left\Vert \vec{t}\right\Vert $
and $t_{\text{fs}}(\vec{t})=-R^{-1}\left\Vert \vec{t}\right\Vert $
. Thus, Eq. \ref{eq:strictConv} reduces to the capacity of balls
is 
\begin{align}
\alpha_{\text{Ball}}^{-1} & =\int_{0}^{\infty}dt\chi_{D}(t)\int_{\kappa-tR^{-1}}^{\kappa+tR}Dt_{0}\frac{(-t_{0}+tR+\kappa)^{2}}{(1+R^{2})}\nonumber \\
 & +\int_{0}^{\infty}dt\chi_{D}(t)\int_{-\infty}^{\kappa-tR^{-1}}Dt_{0}\left[(t_{0}-\kappa)^{2}+t^{2}\right]\label{eq:capacityL2ball}
\end{align}
where, 
\begin{equation}
\chi_{D}(t)=\frac{2^{1-\frac{D}{2}}}{\Gamma(\frac{D}{2})}t^{D-1}e^{-\frac{1}{2}t^{2}},\;t\geq0\label{eq:chi_D-1}
\end{equation}
is the $D$-dimensional Chi probability density function, reproducing
the results of \citep{chung2016linear}. Furthermore, Eq. \ref{eq:capacityL2ball}
can be approximated by the capacity of points with a margin increase
of $R\sqrt{D}$, i.e. $\alpha_{\text{Ball}}\left(\kappa,R_{\text{M}},D_{\text{M}}\right)\approx(1+R^{2})\alpha_{0}(\kappa+R\sqrt{D})$
(details in \citep{chung2016linear}).

\subsection*{B.2. $\ell_{2}$ Ellipsoids\label{sec:AppendixC_l2_ell}}

\subsubsection{Anchor points and support regimes}

With ellipsoids, Eq. \ref{eq:constraint_ellipsoid}, the support function
in Eq. \ref{eq:support_gradient} can be computed explicitly as follows.
For a vector $\vec{V}=(\vec{v},\,v_{0})$, with non zero $\vec{v}$,
the support function $g(\vec{v})$ is minimized by a vector $\vec{s}$
which occurs on the boundary of the ellipsoid, i.e. it obeys the equality
constraint $f(\vec{s})=0$ with $\rho(\vec{s})=\sum_{i=1}^{D}\left(\frac{s_{i}}{R_{i}}\right)^{2}-1$
, Eq. \eqref{eq:constraint_ellipsoid}. To evaluate $g$, we differentiate
$-\sum_{i=1}^{D}s_{i}\vec{v}_{i}+\rho f(\vec{s}$) with respect to
$s_{i}$, where $\rho$ is a Lagrange multiplier enforcing the constraint,
yielding

\begin{equation}
\tilde{s}_{i}=\frac{v_{i}R_{i}^{2}}{g(\vec{v})}\label{eq:stilde-1}
\end{equation}

and

\begin{equation}
g(\vec{v})=-\left\Vert \vec{v}\circ\vec{R}\right\Vert \label{eq:minS-1}
\end{equation}
where $\vec{R}$ is the $D$ dimensional vector of the ellipsoid principal
radii, and we denote $\circ$refers to pointwise product of vectors,
$(\vec{v}\circ\vec{R})_{i}=v_{i}R_{i}$. For a given $(\vec{t},\,t_{0})$,
the vector $(\vec{v},\,v_{0})$ is determined by Eq. \eqref{eq:vlambdagrad}
and the analytic solution above can be used to derive explicit expressions
for $\tilde{s}(\vec{t},t_{0})$ in the different regimes as follows.

\textbf{Interior regime:}

In the interior regime $\lambda=0$ , $\vec{v}=\vec{t}$, resulting
in zero contribution to the inverse capacity. The anchor point is
given by the following boundary point on the ellipse, given by Eq.
\ref{eq:stilde-1} with $\vec{v}=\vec{t}.$ This regime holds for
$t_{0}$ obeying the inequality $t_{0}-\kappa\geq t_{\text{touch}}(\vec{t})$
with Eq. \ref{eq:tTouch}, yielding 
\begin{equation}
t_{\text{touch}}(\vec{t})=\left\Vert \vec{t}\circ\vec{R}\right\Vert \label{eq:ttouch-1}
\end{equation}

\textbf{Touching regime:} Here the anchor point is given by Eq. \ref{eq:stilde-1}
where $\vec{v}=\vec{t}+\lambda\tilde{s}$. Substituting $t_{i}+\lambda\tilde{s_{i}}$
for $v_{i}$ in the numerator of that equation, and $g(\vec{v})=\kappa-v_{0}=\kappa-t_{0}-\lambda$
, yields,

\begin{equation}
\tilde{s}_{\text{touch},i}(\vec{t},t_{0})=-\frac{t_{i}R_{i}^{2}}{\lambda(1+R_{i}^{2})+(-t_{0}+\kappa)},\label{eq:seTouch}
\end{equation}

where the parameter $\lambda$ is determined by the ellipsoidal constraint,
\begin{equation}
1=\sum_{i=1}^{D}\frac{t_{i}^{2}R_{i}^{2}}{\left[\lambda(1+R_{i}^{2})+(-t_{0}+\kappa)\right]^{2}}.\label{eq:lambdaTouch}
\end{equation}

In this regime, the contribution to the capacity is given by Eq. \ref{eq:flambdas}-\ref{eq:lambdaRectified}
with $\tilde{s}$ in Eq. \eqref{eq:seTouch}.

The touching regime holds for $t_{\text{touch}}>t_{0}-\kappa>t_{\text{fs}}$
, where $\lambda\rightarrow\kappa-t_{0}$ , $\vec{v}$ vanishes, and
the anchor point is the point on the boundary of the ellipsoid antiparallel
to $\vec{t}$ . Substituting this value of $\lambda$ in Eq. \ref{eq:lambdaTouch}
, yields, 
\begin{equation}
t_{\text{fs}}(\vec{t})=-\sqrt{\sum_{i}\left(\frac{t_{i}}{R_{i}}\right)^{2}}\label{eq:tembed-1}
\end{equation}

\textbf{Fully supporting regime:} When $t_{0}-\kappa<t_{\text{fs}}$
, we have $\vec{v}_{\text{\ensuremath{}}}=0$, $v_{0}=\kappa,$ and
$\lambda=t_{0}-\kappa$, implying that the center as well as the entire
ellipsoid is fully supporting the max margin solution. In this case,
the anchor point is antiparallel to $\vec{t}$ at the interior point,
\ref{eq:svsVT}, and its contribution to the capacity is as in Eq.
\ref{eq:F_full Supp}.

\section{Limit of Large Manifolds\label{sec:AppendixC_LargeLimit}}

In the large size limit, $\left\Vert \tilde{s}\right\Vert \rightarrow\infty$
, linear separation of manifolds reduces to linear separation of $P$
random $D$-dimensional affine subspaces. When separating subspaces,
all of them must be fully embedded in the margin plane, otherwise
they would intersect it and violate the classification constraints.
However, the way large size manifolds approach this limit is subtle.
To analyze this limit, we note that when $\left\Vert \tilde{s}\right\Vert $
is large, $g(\vec{v})>\left\Vert \tilde{s}\right\Vert \left\Vert \vec{v}\right\Vert $
and from the condition that $g(\vec{v})\geq(-v_{0}+\kappa)$, we have
$\left\Vert \vec{v}\right\Vert \le\frac{\left(-v_{0}+\kappa\right)}{\left\Vert \tilde{s}\right\Vert }$,
i.e., $\left\Vert \vec{v}\right\Vert $ is $O(\left\Vert \tilde{s}\right\Vert ^{-1})$.
A small $\left\Vert \vec{v}\right\Vert $ implies that the affine
basis vectors, except the center direction, are all either exactly
or almost orthogonal to the solution weight vector. Since $\lambda\tilde{s}=-\vec{t}+\vec{v}\approx-\vec{t}$,
it follows that $\tilde{s}$ is almost antiparallel to the gaussian
vector $\vec{t}$, hence $D_{\text{M}}\rightarrow D$, see Eq. \ref{eq:DM}.
To elucidate the manifold support structure, we note first that by
Eq. \ref{eq:tTouch} $t_{\text{touch}}\propto-||\vec{t}||||\tilde{s}||\rightarrow-\infty$
hence the fractional volume of the interior regime is negligible,
and the statistics is dominated by the embedded regimes. In fact,
the fully embedded transition is given by $t_{\text{embed}}\approx-\kappa$
, see Eq.\ref{eq:tembedGen}, so that the fractional volume of the
fully embedded regime is $H(-\kappa)=\int_{-\kappa}^{\infty}Dt_{0}$
and its contribution to inverse capacity is therefore $\int_{-\kappa}^{\infty}Dt_{0}\left[\langle\left\Vert \vec{t}\right\Vert ^{2}\rangle+(t_{0}+\kappa)^{2}\right]=H(-\kappa)D+\alpha_{0}^{-1}(\kappa)$
. The remaining summed probability of the touching and partially embedded
regimes ($k\geq1)$ is therefore $H(\kappa$). In these regimes, $\lambda^{2}\left\Vert \tilde{s}\right\Vert ^{2}\approx\lambda^{2}\left\Vert \tilde{s}\right\Vert ^{2}\approx\left\Vert \vec{t}\right\Vert ^{2}$
, so that this regime contributes a factor of $\int_{-\infty}^{-\kappa}Dt_{0}\langle\left\Vert \vec{t}\right\Vert ^{2}\rangle=H(\kappa)D$.
Combining these two contributions, we obtain for large sizes, $\alpha_{\text{M}}^{-1}=D+\alpha_{0}^{-1}(\kappa)$,
consistent with Eq. \eqref{eq:PNboundGardner}.

\section{High Dimensional Manifolds\label{sec:AppendixB_highD-1}}

\subsection*{D.1. High Dimensional $\ell_{2}$ ball}

Before we discuss general manifolds in high dimension, we focus on
the simple case of high dimensional balls, the capacity of which is
given by Eq. \ref{eq:capacityL2ball}. When $D\gg1$ , $\chi_{D}(t)$
, \ref{eq:chi_D-1}, is centered around $t=\sqrt{D}$. Substituting
$\chi_{D}(t)\approx\delta(t-\sqrt{D}$) yields

\begin{equation}
\alpha_{\text{Ball}}^{-1}(\kappa,R,D)=\int_{\kappa-\frac{\sqrt{D}}{R}}^{\kappa+R\sqrt{D}}Dt_{0}\frac{(R\sqrt{D}+\kappa-t)_{0}^{2}}{R^{2}+1}+\int_{-\infty}^{\kappa-\frac{\sqrt{D}}{R}}Dt_{0}([t_{0}-\kappa]^{2}+D)\label{eq:alphacSpheresLargeD}
\end{equation}
As long as $R\ll\sqrt{D}$ , the second term in Eq. (\ref{eq:alphacSpheresLargeD})
vanishes and yields

\begin{equation}
\alpha_{\text{Ball}}^{-1}(\kappa,R,D)=\int_{-\infty}^{\kappa+R\sqrt{D}}Dt_{0}\frac{(R\sqrt{D}+\kappa-t_{0})^{2}}{R^{2}+1}\label{eq:scalingAlpha}
\end{equation}

Here we note that the $t_{0}$ term in the numerator is significant
only if $R=O(D^{-1/2})$ or smaller, in which case the denominator
is just $1$ . When $R$ is of order $1$ the term $t_{0}$ in the
integrand is negligible. Hence, in both cases we can write,

\begin{equation}
\alpha_{\text{Ball}}^{-1}(\kappa,R,D)\approx\alpha_{0}^{-1}\left(\frac{\kappa+R\sqrt{D}}{\sqrt{1+R^{2}}}\right)\label{eq:ball_point}
\end{equation}
In this form, the intuition beyond the factor $\sqrt{1+R^{2}}$ is
clear, stemming from the fact that the distance of a point from the
margin plane scales with its norm, hence the margin entering the capacity
should factor out this norm.

As stated above, Eq, \ref{eq:ball_point} implies a finite capacity
only in the scaling regime, where $R\sqrt{D}=O(1).$ If, on the other
hand, $R\sqrt{D}\gg1$, Eq. (\ref{eq:scalingAlpha}) implies

\begin{equation}
\alpha_{\text{Ball}}^{-1}=\frac{R^{2}D}{1+R^{2}}\label{eq:alphaD}
\end{equation}
(where we have used the asymptote $\alpha_{0}^{-1}(x)\rightarrow x^{2}$
for large $x$), reducing to $\alpha^{-1}=D$ for large $R$.

The analysis above also highlights the support structure of the balls
at large $D.$ As long as $R\ll\sqrt{D}$ the

fraction of balls that lie fully in the margin plane, is negligible,
as implied by the fact that $t_{fs}\approx-\sqrt{D}/R\rightarrow-\infty$.The
overall fraction of interior balls is $H(\kappa+R\sqrt{D})$ whereas
the fraction that touch the margin planes is $1-H(\kappa+R\sqrt{D})$
. Despite the fact that there are no fully supporting balls, the touching
balls are almost parallel to the margin planes if $R\gg1$ hence the
capacity reaches its lower bound.

Finally, the large manifold limit discussed in Appendix \ref{sec:AppendixC_LargeLimit}
is realized for as $R\ll\sqrt{D}$ . Here, $t_{fs}=-\kappa$ and the
system is either touching or fully supporting with probabilities $H(\kappa)$
and $H(-\kappa)$, respectively.

\subsection*{D.2. General manifolds. }

To analyze the limit of high dimensional general manifolds we utilize
the self averaging of terms in Eq. \eqref{eq:amMgen}, involving sums
of the $D$ components of $\vec{t}$ and $\tilde{s}$, i.e., 
\begin{align}
\vec{t}\cdot\tilde{s} & \approx\langle||\tilde{s}||\rangle\langle\vec{t}\cdot\hat{s}\rangle\approx R_{\text{M}}\sqrt{D_{\text{M}}}=\kappa_{\text{M}}\nonumber \\
\label{eq:highDSelfAverage1-1}\\
\nonumber 
\end{align}

Also, $t_{\text{touch}}\approx\vec{t}\cdot\tilde{s}\approx\kappa_{\text{M}}$
, hence, we obtain for the capacity, 
\begin{align}
\alpha_{\text{M}}^{-1} & (\kappa)\approx\frac{\langle[\kappa_{\text{M}}+\kappa-t_{0}]_{+}^{2}\rangle_{t_{0}}}{R_{\text{M}}^{2}+1}\nonumber \\
 & \approx\alpha_{\text{Ball}}^{-1}(\kappa,R_{\text{M}},D_{\text{M}})\label{eq:capacityLargeD-1-1-1-1}
\end{align}

where the average is wrt the gaussian $t_{0}$. Evaluating $R_{\text{M}}$
and $D_{\text{M}}$ involves calculations of the self consistent statistics
of the anchor points. This calculation is simplified in high dimension.
In particular, $\lambda$, Eq. \ref{eq:lambdaRectified} , reduces
to 
\begin{equation}
\lambda\approx\frac{\langle[\kappa_{\text{M}}+\kappa-t_{0}]_{+}\rangle_{t_{0}}}{1+R_{\text{M}}^{2}}\label{eq:lambdaHighD-1}
\end{equation}

hence it is approximately constant, independent of $\vec{T}$.

In deriving the above approximations we used self-averaging in summations
involving the $D$ intrinsic coordinates. The full dependence on the
longitudinal gaussian $t_{0}$ should remain. Thus, $R_{\text{M}}$and
$D_{\text{M}}$ should in fact be substituted by $R_{\text{M}}(t_{0})$
and $D_{\text{M}}(t_{0})$ denoting \ref{eq:RM}-\ref{eq:DM} with
averaging only over $\vec{t}$. This would yield a more complicated
expressions than Eq.\ref{eq:capacityLargeD-1-1-1-1}- \ref{eq:lambdaHighD-1}.

The reason why we can replace them by average quantities is the following:
the potential dependence of the anchor radius and dimension on $t_{0}$
is via $\lambda(t_{0}).$ However, inspecting Eq. \ref{eq:lambdaHighD-1}
we note two scenarios: One in which $R_{\text{M}}$ is small and $\kappa_{\text{M}}$
of order $1$ . In this case, because the manifold radius is small
the contribution $\lambda\tilde{s}$ is small and can be neglected.
This is the same argument why in this case, the geometry can be replaced
by the Gaussian geometry which does not depend on $t_{0}$ or $\kappa$
. The second scenario is that $R_{\text{M}}$ is of order $1$ and
$\kappa_{\text{M}}\gg1$, in which case the order $1$ contribution
from $t_{0}$ is negligible.

\section{Capacity of $\ell_{2}$ Balls with Sparse Labels\label{sec:AppendixBsparse}}

First, we note that the capacity of sparsely labeled points is $\alpha_{0}(f,\kappa)=\mbox{max}_{b}\alpha_{0}(f,\kappa,b)$,
where

\begin{align}
\alpha_{0}^{-1}(f,\kappa,b) & =f\int_{-\infty}^{\kappa+b}Dt(-t+\kappa+b)^{2}\nonumber \\
 & +(1-f)\int_{-\infty}^{\kappa-b}Dt(-t+\kappa-b)^{2}
\end{align}

Optimizing $b$ yields the following equation for $b$, 
\begin{equation}
0=f\int_{-\infty}^{b+\kappa}Dt(-t+\kappa+b)+(1-f)\int_{-\infty}^{b-\kappa}Dt(-t+\kappa-b)
\end{equation}

In the limit of $f\rightarrow0$, $b\gg1$ . The first equation reduces
to

\begin{equation}
\alpha_{0}^{-1}\approx fb^{2}+\exp\left(-b^{2}/2\right)
\end{equation}

yielding for the optimal $b\approx\sqrt{2|\log f|}$ . With this $b$,
the inverse capacity is dominated by the first term which is $\alpha_{0}^{-1}\approx2f|\log f|$
.

The capacity for $\ell_{2}$ balls of radius $R$ with sparse labels
with sparsity $f$ (Eq. \eqref{eq:alpha_sparse_M} and Eq. \ref{eq:capacityL2ball})
is given by, 
\begin{align}
\alpha_{\text{Ball}}^{-1} & =f\int_{0}^{\infty}dt\chi_{D}(t)\text{\ensuremath{\cdot}}\nonumber \\
 & \left[\int_{\kappa-tR^{-1}+b}^{\kappa+tR+b}Dt_{0}\frac{(-t_{0}+tR+b+\kappa)^{2}}{(1+R^{2})}\right.\nonumber \\
 & \left.+\int_{-\infty}^{b+\kappa-tR^{-1}}Dt_{0}((t_{0}-b-\kappa)^{2}+t^{2})\right]\nonumber \\
 & +(1-f)\int_{0}^{\infty}dt\chi_{D}(t)\text{x}\nonumber \\
 & \left[\int_{\kappa-tR^{-1}-b}^{\kappa+tR-b}Dt_{0}\frac{(-t_{0}+tR+\kappa-b)^{2}}{(1+R^{2})}\right.\nonumber \\
 & \left.+\int_{-\infty}^{-b+\kappa-tR^{-1}}Dt_{0}((t_{0}-\kappa+b)^{2}+t^{2})\right]\label{eq:Sparse_alphaB}
\end{align}
and the optimal bias $b$ is given by $\partial\alpha_{Ball}/\partial b=0$.
Here we analyze these equations in various size regimes assuming $\kappa=0$.

\textbf{Small $R$:} For balls with small radius, the capacity is
that of points, unless the dimensionality is high. Thus, if $D$ is
large but $R\sqrt{D}\lesssim1$,

\begin{align}
\alpha_{\text{Ball}}^{-1} & =f\int_{-\infty}^{R\sqrt{D}+b}Dt_{0}\frac{(-t_{0}+R\sqrt{D}+b)^{2}}{(1+R^{2})}\nonumber \\
 & +(1-f)\int_{-\infty}^{R\sqrt{D}-b}Dt_{0}\frac{(-t_{0}+R\sqrt{D}+b)^{2}}{(1+R^{2})}\nonumber \\
\nonumber \\
\nonumber \\
\nonumber \\
\label{eq:Sparse_alphaB-1}
\end{align}

that is, 
\begin{equation}
\alpha_{\text{Ball}}(f,R,D)\approx\alpha_{0}(f,\kappa=R\sqrt{D})
\end{equation}

As noted above when $f\rightarrow0$ the optimal bias diverges, hence
the presence of order $1$ the induced margin is noticeable only for
moderate $f$ such that $R\sqrt{D}=O(|\log f|).$

\textbf{Small $f$ and large $R$:} Here we analyze the above equations
in the limit of small $f$ and large $R$. We assume that $f$ is
sufficiently small so that the optimal bias $b$ is large. The contribution
of the minority class to the inverse capacity $\alpha_{Ball}^{-1}$
is dominated by

\begin{equation}
f\int_{0}^{\infty}dt\chi_{D}(t)\int_{-\infty}^{b}Dt_{0}((-t_{0}+b)^{2}+t^{2})\approx fb^{2}\label{eq:SparseL2_smallfLargeR_f}
\end{equation}
The dominant contribution of the majority class to $\alpha_{Ball}^{-1}$
is,

\begin{align}
 & (1-f)\int_{0}^{\infty}dt\chi_{D}(t)\int_{-b-Rt^{-1}}^{-b+tR}Dt_{0}\frac{(-t_{0}+tR-b)^{2}}{(1+R^{2})}\\
 & \approx(1-f)\int_{\bar{b}}^{\infty}dt\chi_{D}(t)(t-\bar{b})^{2}\label{eq:SparseL2_smallfLargeR_1-f}
\end{align}

where $\bar{b}=b/R$ . In deriving the last equation we used $\frac{(-t_{0}+tR-b)^{2}}{(1+R^{2})}\rightarrow(t-\bar{b})^{2}$as
$R,b\rightarrow\infty$ . Second, the integrals are of substantial
value only if $t\geq\bar{b}$ in which case the integral of $t_{0}$
is $\int_{-R(\bar{b}-t)}^{b}Dt_{0}\approx1$ and the integral over
$t$ is from $\bar{b}$ to $\infty$. Combining the two results yields
the following simple expression for the inverse capacity,

\begin{equation}
\alpha_{\text{Ball}}^{-1}=\bar{f}\bar{b}^{2}+\int_{\bar{b}}^{\infty}dt\chi_{D}(t)(t-\bar{b})^{2}\label{eq:alfbarx}
\end{equation}

where scaled sparsity is $\bar{f}=fR^{2}$. The optimal scaled bias
is given by,

\begin{equation}
\bar{f}\bar{b}=\int_{\bar{b}}^{\infty}dt\chi_{D}(t)(t-\bar{b})\label{eq:eqforx}
\end{equation}

Note that $R$ and $f$ affect the capacity only through the scaled
sparsity. When $\bar{f}\rightarrow0$, capacity is proportional to
$(\bar{f}|\log\bar{f}|)^{-1}$. In a realistic regime of small $\bar{f}$
(between $10^{-4}$to $1$) capacity decreases with $\bar{f}$ roughly
as $1/\bar{f}$ with a proportionality constant which depends on $D$,
as shown in the examples of Fig. 12 (a). Finally, when $R$ is sufficiently
large that $\bar{f}>1$, $\bar{b}$ is order 1 or smaller. In this
case, the second term in Eq. \ref{eq:alfbarx} dominates and contributes
$\langle t^{2}\rangle=D$ yielding a capacity that saturates as $D^{-1}$
as shown in Fig. \ref{fig:SparseManifolds}(a). It should be noted
however, that when $b$ is not large, the approximations Eq. \ref{eq:alfbarx}-\ref{eq:eqforx}
don't hold anymore.

\bibliographystyle{unsrt}
\bibliography{prx_genmanifolds}

\end{document}